\documentclass[twocolumn,iop,numberedappendix,appendixfloats]{openjournal}
\makeatletter
\bibpunct{[}{]}{,}{n}{}{,} 
\makeatother
\usepackage[utf8]{inputenc}
\usepackage{amsmath,amssymb,amstext}
\usepackage{graphicx}
\usepackage{xcolor}
\graphicspath{ {./images/} }
\usepackage{booktabs}
\usepackage{array}
\usepackage{bm}
\usepackage{subcaption}
\usepackage[colorlinks,allcolors=blue]{hyperref}
\usepackage[capitalise,nameinlink]{cleveref}
\usepackage{dsfont}
\usepackage{comment}
\usepackage{pifont}
\usepackage{listings}
\usepackage{fontawesome}
\usepackage[normalem]{ulem}
\definecolor{maroon}{cmyk}{0, 0.87, 0.68, 0.32}
\definecolor{halfgray}{gray}{0.55}
\definecolor{ipython_frame}{RGB}{207, 207, 207}
\definecolor{ipython_bg}{RGB}{247, 247, 247}
\definecolor{ipython_red}{RGB}{186, 33, 33}
\definecolor{ipython_green}{RGB}{0, 128, 0}
\definecolor{ipython_cyan}{RGB}{64, 128, 128}
\definecolor{ipython_purple}{RGB}{170, 34, 255}

\newcommand{\de}{\mathrm{d}}
\newcommand{\pd}{\partial}
\newcommand{\fnl}{\ensuremath{f_{\rm NL}}}

\lstdefinelanguage{iPython}{
    morekeywords={access,and,break,class,continue,def,del,elif,else,except,exec,finally,for,from,global,if,import,in,is,lambda,not,or,pass,print,raise,return,try,while},%
    morekeywords=[2]{abs,all,any,basestring,bin,bool,bytearray,callable,chr,classmethod,cmp,compile,complex,delattr,dict,dir,divmod,enumerate,eval,execfile,file,filter,float,format,frozenset,getattr,globals,hasattr,hash,help,hex,id,input,int,isinstance,issubclass,iter,len,list,locals,long,map,max,memoryview,min,next,object,oct,open,ord,pow,property,range,raw_input,reduce,reload,repr,reversed,round,set,setattr,slice,sorted,staticmethod,str,sum,super,tuple,type,unichr,unicode,vars,xrange,zip,apply,buffer,coerce,intern, function, @model, Uniform, Normal, MvNormal, theory_planck},%
    sensitive=true,%
    morecomment=[l]\#,%
    morestring=[b]',%
    morestring=[b]",%
    morestring=[s]{'''}{'''},
    morestring=[s]{"""}{"""},
    morestring=[s]{r'}{'},
    morestring=[s]{r"}{"},%
    morestring=[s]{r'''}{'''},%
    morestring=[s]{r"""}{"""},%
    morestring=[s]{u'}{'},
    morestring=[s]{u"}{"},%
    morestring=[s]{u'''}{'''},%
    morestring=[s]{u"""}{"""},%
    literate=
    {á}{{\'a}}1 {é}{{\'e}}1 {í}{{\'i}}1 {ó}{{\'o}}1 {ú}{{\'u}}1
    {Á}{{\'A}}1 {É}{{\'E}}1 {Í}{{\'I}}1 {Ó}{{\'O}}1 {Ú}{{\'U}}1
    {à}{{\`a}}1 {è}{{\`e}}1 {ì}{{\`i}}1 {ò}{{\`o}}1 {ù}{{\`u}}1
    {À}{{\`A}}1 {È}{{\'E}}1 {Ì}{{\`I}}1 {Ò}{{\`O}}1 {Ù}{{\`U}}1
    {ä}{{\"a}}1 {ë}{{\"e}}1 {ï}{{\"i}}1 {ö}{{\"o}}1 {ü}{{\"u}}1
    {Ä}{{\"A}}1 {Ë}{{\"E}}1 {Ï}{{\"I}}1 {Ö}{{\"O}}1 {Ü}{{\"U}}1
    {â}{{\^a}}1 {ê}{{\^e}}1 {î}{{\^i}}1 {ô}{{\^o}}1 {û}{{\^u}}1
    {Â}{{\^A}}1 {Ê}{{\^E}}1 {Î}{{\^I}}1 {Ô}{{\^O}}1 {Û}{{\^U}}1
    {œ}{{\oe}}1 {Œ}{{\OE}}1 {æ}{{\ae}}1 {Æ}{{\AE}}1 {ß}{{\ss}}1
    {ç}{{\c c}}1 {Ç}{{\c C}}1 {ø}{{\o}}1 {å}{{\r a}}1 {Å}{{\r A}}1
    {€}{{\EUR}}1 {£}{{\pounds}}1
    {^}{{{\color{ipython_purple}\^{}}}}1
    {=}{{{\color{ipython_purple}=}}}1
    {+}{{{\color{ipython_purple}+}}}1
    {-}{{{\color{ipython_purple}-}}}1
    {*}{{{\color{ipython_purple}$^\ast$}}}1
    {/}{{{\color{ipython_purple}/}}}1
    {+=}{{{+=}}}1
    {-=}{{{-=}}}1
    {*=}{{{$^\ast$=}}}1
    {/=}{{{/=}}}1,
    literate=
    *{-}{{{\color{ipython_purple}-}}}1
     {?}{{{\color{ipython_purple}?}}}1,
    identifierstyle=\color{black}\ttfamily,
    commentstyle=\color{ipython_cyan}\ttfamily,
    stringstyle=\color{ipython_red}\ttfamily,
    keepspaces=true,
    showspaces=false,
    showstringspaces=false,
    rulecolor=\color{ipython_frame},
    frameround={t}{t}{t}{t},
    numbers=none,
    numberstyle=\tiny\color{halfgray},
    backgroundcolor=\color{ipython_bg},
    basicstyle=\ttfamily\footnotesize,
    columns=fullflexible,
    keywordstyle=\color{ipython_green}\ttfamily,
}

\newcommand{\blast}{\texttt{Blast.jl}}
\newcommand{\swift}{\texttt{SwiftC}$_\ell$}
\newcommand{\threebytwo}{$3\times2$ pt}
\newcommand{\julia}{\texttt{Julia}}
\newcommand{\nk}{\texttt{N5K}}

\date{\today}

\begin{document}
\journalinfo{The Open Journal of Astrophysics}
\submitted{submitted September 2026; accepted Oct 2026}

\shorttitle{\blast{}: Differentiable Non-Limber Power Spectra}
\shortauthors{Chiarenza et al.}
\title{\blast{}: Differentiable Non-Limber Power Spectra for Joint Clustering, Shear, and CMB lensing Analyses}
\author{Sofia Chiarenza$^{\star1,2,3}$}
\author{Marco Bonici$^{1,2,3}$}
\author{Stefano Camera$^{4,5,6,7}$}
\author{Giulio Fabbian$^{8}$}
\author{Carlos García-García$^{1,2,9}$}
\author{Alex Krolewski$^{1,2,10}$}
\author{Will J.\ Percival$^{1,2,3}$}

\affiliation{$^1$ Waterloo Centre for Astrophysics, University of Waterloo, Waterloo, ON N2L 3G1, Canada}
\affiliation{$^2$ Department of Physics and Astronomy, University of Waterloo, Waterloo, ON N2L 3G1, Canada}
\affiliation{$^3$ Perimeter Institute for Theoretical Physics, 31 Caroline St North, Waterloo, ON N2L 2Y5, Canada}
\affiliation{$^4$ Dipartimento di Fisica, Universit\`a degli Studi di Torino, Via P.\ Giuria 1, 10125 Torino, Italy}
\affiliation{$^5$ INFN -- Istituto Nazionale di Fisica Nucleare, Sezione di Torino, Via P.\ Giuria 1, 10125 Torino, Italy}
\affiliation{$^6$ INAF -- Istituto Nazionale di Astrofisica, Osservatorio Astrofisico di Torino, Strada Osservatorio 20, 10025 Pino Torinese, Italy}
\affiliation{$^7$ Department of Physics \& Astronomy, University of the Western Cape, Cape Town 7535, South Africa}
\affiliation{$^8$ Universit\'e Paris-Saclay, CNRS, Institut d'astrophysique spatiale, 91405, Orsay, France}
\affiliation{$^9$ CIEMAT, Avenida Complutense 40, E-28040 Madrid, Spain}
\affiliation{$^{10}$ California Institute of Technology, 1200~East California Boulevard, Pasadena, CA 91125, USA}

\thanks{$^\star$ E-mail: \url{schiaren@uwaterloo.ca}}

\begin{abstract}
Upcoming large-scale structure surveys require accurate theoretical predictions for angular power spectra on the largest angular scales, where the commonly used Limber approximation breaks down and multiple relativistic and observational effects become relevant. At the same time, modern cosmological analyses increasingly rely on gradient-based inference techniques, motivating the development of fast, fully differentiable algorithms.
In this work, we present a major extension of \blast{}, a numerical toolkit for the efficient computation of non-Limber angular power spectra. Building on the Chebyshev-polynomial decomposition of the original framework, the updated algorithm incorporates redshift-space distortions, magnification bias, primordial non-Gaussianity, intrinsic alignments, CMB lensing, and the integrated Sachs-Wolfe effect, while remaining fully differentiable through custom automatic differentiation rules. Despite the increased physical complexity, the algorithm retains favorable scaling by precomputing all cosmology-independent quantities. We validate the expanded framework against brute-force integration and established cosmological codes, and provide a detailed analysis of the trade-off between computational speed and precision as a function of the algorithm hyperparameters.
We demonstrate the readiness of \blast{} for gradient-based inference through a simulated likelihood analysis with LSST Y10-like mock data, recovering unbiased cosmological and nuisance parameter constraints from a $35$-parameter model using gradient-based samplers.
\end{abstract}

\keywords{%
Cosmology: \threebytwo{}  statistics, non-Limber angular power spectra-- Methods: statistical, data analysis
}
\maketitle
\section{Introduction}\label{sec:introduction}
The next generation of ``Stage-IV'' cosmological surveys, including the Dark Energy Spectroscopic Instrument (DESI) \citep{aghamousa2016desi}, the Vera C. Rubin Observatory's Legacy Survey of Space and Time (LSST) \citep{ivezic2019lsst}, and the Euclid \citep{mellier2025euclid}, Roman \citep{spergel2015wide}, and SPHEREx \citep{dore2014cosmology} space missions, will provide high-precision measurements of the large-scale structure (LSS) of the Universe with unprecedented statistical power. While each of these experiments is remarkably powerful on its own, fully exploiting their information content requires a multi-probe framework that combines different observables. 

By jointly analyzing galaxy clustering, weak gravitational lensing, and CMB observables, one can simultaneously exploit three complementary advantages over single-probe analyses: access to new physics encoded in cross-correlations between different tracers of the matter distribution, improved robustness to survey-specific systematic uncertainties through internal consistency checks across probes, and the ability to break parameter degeneracies therefore improving the constraining power. The CMB contributes two distinct probes: CMB lensing convergence, which traces the projected matter distribution along the line of sight, and the integrated Sachs-Wolfe (ISW) effect, detected through its cross-correlation with galaxy surveys. The ISW effect is sensitive to the time evolution of gravitational potentials during the epoch of accelerated expansion, making it a powerful probe of dark energy and modifications of gravity \citep{Renk:2017rzu, Stolzner:2017ged, Krolewski:2021znk, seraille2024constraining, ghodsi2026probing, garcia2026status}. 
Together, these advantages enable qualitatively new constraints on cosmological models \citep{di2021realm, abdalla2022cosmology}, including tighter bounds on parameters such as $S_8$ and $H_0$ where single-probe analyses currently show discrepancies \citep{abbott2022dark, hikage2019cosmology, garcia2021growth, heymans2021kids, singh2017cross, doux2018cosmological, krolewski2020unwise, sailer2025cosmological, qu2025atacama, shaikh2024cosmology, troster2022joint, Garcia-Garcia:2024gzy}.

Implementing such a joint analysis at the required level of precision poses significant theoretical and computational challenges. It demands accurate predictions for angular power spectra across a broad range of scales, including the largest angular scales where the traditionally used Limber approximation \citep{limber1953analysis} is no longer sufficient.
In fact, angular power spectra of projected cosmological observables are defined by integrals over spherical Bessel functions, which are highly oscillatory and numerically challenging to evaluate. For this reason, most current analyses \citep{secco2022dark, asgari2021kids, frusciante2025euclid} rely on the Limber or extended-Limber approximations \citep{limber1953analysis, loverde2008extended}, which reduce the dimensionality of the problem and allow for fast numerical evaluation. While these approximations are accurate at sufficiently high multipoles, they break down on large angular scales and for narrow redshift kernels. This regime is precisely where several effects of interest such as primordial non-Gaussianity, redshift-space distortions, and cross-correlations with CMB probes become most relevant.

The \blast{} algorithm \citep{chiarenza2024blast} was introduced to address this challenge by providing an efficient and accurate method for computing beyond-Limber angular power spectra. Its key idea is to decompose the three-dimensional matter power spectrum onto a Chebyshev polynomial basis, isolating the highly oscillatory integrals into a set of cosmology-independent quantities that can be precomputed and reused. To ensure high performance and flexibility, \blast{} was originally implemented as a \julia{} package. In its initial formulation, the algorithm demonstrated an order-of-magnitude improvement in computational efficiency compared to existing beyond-Limber methods, highlighting the effectiveness of the underlying decomposition strategy.

Contemporary and upcoming analyses require the joint modeling of multiple probes and a broader set of physical effects, extending beyond galaxy number counts and cosmic shear. In this work, we expand the physical scope of the existing observables to include redshift-space distortions (RSD), magnification bias, primordial non-Gaussianity (PNG), and intrinsic alignments (IA), and broaden the set of supported probes to encompass cross-correlations with CMB observables, namely CMB lensing and the integrated Sachs-Wolfe (ISW) effect. These additions are numerically non-trivial: RSD introduces second derivatives of spherical Bessel functions, PNG requires handling multiple classes of unequal-time power spectra due to its scale-dependent bias, and the new CMB kernels extend the range of projection functions that must be mapped onto the precomputed basis integrals.

The computation of derivatives is ubiquitous in cosmological data analysis. Derivatives of theoretical predictions and likelihoods with respect to cosmological and nuisance parameters are required in a variety of applications, ranging from Fisher-matrix forecasts, which rely on Jacobians of the predicted observables or on the curvature of the likelihood, to numerical optimization and modern gradient-based sampling methods \citep{Euclid:2019clj, Euclid:2020fjs, Euclid:2022hdx, paganin2024euclid, Sarcevic:2024tdr}. Traditionally, cosmological analyses have often relied on finite-difference techniques to evaluate these derivatives. While straightforward to implement, finite differences require the choice of an appropriate step size, can be affected by the competing effects of truncation and numerical errors, and often require dedicated algorithms to ensure the robustness of the computations \citep{Camera:2016owj, vsarvcevic2026derivkit}. A powerful alternative, originating from developments in computer science and numerical optimization, is automatic differentiation (AD) \citep{griewank2003mathematical}, which systematically applies the chain rule to the operations composing a numerical program and provides derivatives without the step-size tuning required by finite differences.

To make \blast{} compatible with this broader class of derivative-based applications, we have implemented custom gradient rules for AD. Building on the native \julia{} architecture, these hand-optimized rules maximize the performance of the differentiable pipeline and enable efficient derivatives with respect to both cosmological and nuisance parameters. In particular, this makes \blast{} well suited for high-dimensional gradient-based inference methods such as Hamiltonian Monte Carlo \citep{neal2011mcmc}, while also enabling applications such as Fisher forecasting and gradient-based optimization. This paper, which described these changes and their testing is organized as follows: Sec.~\ref{sec:original_blast} briefly 
summarizes the original \blast{} algorithm for completeness. Sec.~\ref{sec:extensions} describes the extensions to the \blast{} algorithm required to support the full framework, covering the generalized power spectrum treatment, the expanded numerical infrastructure, and the projection kernels for each physical observable. Sec.~\ref{sec:performance} presents performance benchmarks and validation of the extended framework, including comparisons against brute-force integration and well-established codes. Sec.~\ref{sec:likelihood} demonstrates the readiness of \blast{} for realistic inference through a simulated likelihood analysis with LSST Y10 mock data. Finally, Sec.~\ref{sec:conclusions} summarizes our results and discusses future directions for the \blast{} framework.

\section{The original \blast{} algorithm}\label{sec:original_blast}
We summarize the \blast{}\footnote{\url{https://github.com/sofiachiarenza/Blast.jl}} algorithm here for completeness, a comprehensive and detailed  derivation is provided in \citep{chiarenza2024blast}. 
The original implementation allows for the computation of angular power spectra for galaxy number counts, cosmic shear, and their cross-correlations without the use of the Limber approximation. This involves the evaluation of a double line-of-sight integral over a third $k$-integral containing highly-oscillating spherical Bessel functions $j_\ell$. These integrals take the form:
\begin{multline}\label{eqn:cls-general}
    C_{ij,\ell}^{\mathrm{AB}} = N^{\mathrm{AB}}_\ell \, \int_0^\infty \de  \chi_1 \, W_i^{\mathrm{A}}(\chi_1) \,
    \int_0^\infty \de  \chi_2 \, W_j^{\mathrm{B}}(\chi_2) \\
    \times \int_0^\infty \de k \, k^2\,P_{\mathrm{AB}}(k,\chi_1, \chi_2)\,
    \frac{j_\ell(k\,\chi_1)}{(k\,\chi_1)^\alpha}\,
    \frac{j_\ell(k\,\chi_2)}{(k\,\chi_2)^\beta}\;.
\end{multline}
where $W_{i,j}^{\mathrm{AB}}(z)$ are window functions: they are survey-specific objects that depend on $n_i(z)$, the redshift distribution of tracer $i$, and the background cosmology. They account for the fact that a real survey measures number counts and cosmic shear in different tomographic redshift bins. The exponents $\alpha$ and $\beta$ are determined by the specific probe (e.g.\ $0$ for density tracers and $2$ for lensing). Due to the nature of the spherical Bessel functions, which are highly oscillating and slowly damped, the evaluation of the integral in Eq.~\ref{eqn:cls-general} is computationally intensive. 

The key idea of \blast{} is to isolate this numerical bottleneck by decomposing the three-dimensional power spectrum into a Chebyshev polynomial basis \citep{trefethen2019approximation},
\begin{equation}
P_{\mathrm{AB}}(k,\chi_1,\chi_2;\bm\theta) \approx \sum_{n=0}^{n_{\mathrm{cheb}}} c_n^{\mathrm{AB}}(\chi_1,\chi_2;\bm\theta)\, T_n(k)\;,
\end{equation}
where all dependence on the cosmological parameter set $\bm\theta$ is encoded in the coefficients $c_n^{\mathrm{AB}}$, while the Chebyshev polynomials $T_n(k)$ remain fixed. This decomposition allows the $k$-integral to be rewritten as a sum over precomputed basis integrals
\begin{equation}
w_{ij,\ell}^{\mathrm{AB}}(\chi_1,\chi_2;\bm\theta)=
\sum_{n=0}^{n_{\mathrm{cheb}}} c_n^{\mathrm{AB}}(\chi_1,\chi_2;\bm\theta)\, \tilde T^{\mathrm{AB}}_{n, \ell}(\chi_1,\chi_2),
\end{equation}
where the basis integrals
\begin{equation}
\tilde T^{\mathrm{AB}}_{n,\ell}(\chi_1,\chi_2) \equiv \int_{k_{\min}}^{k_{\max}}
\de k\, f^{\mathrm{AB}}(k)\, T_n(k)\,j_\ell(k\chi_1)\, j_\ell(k\chi_2)
\end{equation}
depend only on the geometry and the type of observable. Here $f^{\mathrm{AB}}(k)$ is a $k$-dependent prefactor encoding the observable type: each density tracer ($\mathrm{g}$) entering the pair $\mathrm{AB}$ contributes a factor of $k$, while each lensing tracer ($\mathrm{s}$) contributes a factor of $1/k$, so that
\begin{equation}
    f^{\mathrm{AB}}(k) = \begin{cases} k^2 & \mathrm{AB} = \mathrm{gg}\,, \\ 1/k^2 & \mathrm{AB} = \mathrm{ss}\,, \\ 1 & \mathrm{AB} = \mathrm{gs}\,. \end{cases}
\end{equation}
The hyperparameter $n_{\mathrm{cheb}}$ determines the order of the expansion and directly controls the numerical accuracy of the integration, while $[k_{\min}, k_{\max}]$ defines the range over which the power spectrum is sampled. Crucially, these basis integrals are cosmology-independent and are precomputed once, removing the primary cost of the $k$-integration from the parameter inference loop. The expansion coefficients $c_n^{\mathrm{AB}}$ are then evaluated efficiently at each step using a Discrete Fourier Transform (DFT) \citep{press2007numerical, frigo1997fastest}. 

The final ingredient enabling an efficient evaluation of the harmonic-space power spectra is a change of integration variables. Defining $R \equiv \chi_2 / \chi_1$, we rewrite the integrals in terms of the variable pair $(\chi, R)$ rather than $(\chi_1, \chi_2)$. This transformation improves the numerical sampling of the region that provides the dominant contribution to the signal, namely configurations with $\chi_1 \simeq \chi_2$, corresponding to $R \simeq 1$. Further details on this coordinate transformation can be found in Appendix~A of \citep{chiarenza2024blast}.  

In these variables, the angular power spectrum can be written as
\begin{multline}\label{eqn:final_cls}
    C_{ij}^{\mathrm{AB}}(\ell) = \int_{0}^{\infty} \de \chi \, \chi\, \int_0^1 \de R\, 
    \bigl[ \mathcal{K}_i^{\mathrm{A}}(\chi)\mathcal{K}_j^{\mathrm{B}}(R\chi) \\
    + \mathcal{K}_j^{\mathrm{B}}(\chi)\mathcal{K}_i^{\mathrm{A}}(R\chi) \bigr] 
    w_\ell^{\mathrm{AB}}(\chi, R\chi)\;.
\end{multline}
Here,
\begin{equation}
    \mathcal{K}_i^{\mathrm{A}}(\chi) =
    \begin{cases}
        K_i^{\mathrm{A}}(\chi) & \text{for clustering}\\
        K_i^{\mathrm{A}}(\chi)/\chi^2 & \text{for lensing.}
    \end{cases}\;
\end{equation} 
For details of the numerical implementation, we refer the reader to Sec.~3.2 of \citep{chiarenza2024blast}. 

\blast{} treats the linear and non-linear components separately, namely
\begin{equation}\label{eq:pk_dec}
    P(k, \chi_1, \chi_2) = P_{\mathrm{lin}}(k, \chi_1, \chi_2) + [P - P_{\mathrm{lin}}](k, \chi_1, \chi_2)\;.
\end{equation}
The linear component, $P_{\mathrm{lin}}$, is evaluated using the full Chebyshev-based non-Limber framework described above. In contrast, the non-linear residual is important only on small scales where the Limber approximation is highly accurate. Following \citep{chisari2019unequal}, who showed that unequal-time non-linear contributions, arising from $\chi_1 \neq \chi_2$, decay exponentially as $\propto \exp\left(-(D(\chi_1) - D(\chi_2))^2\right)$, we correct the angular power spectrum for non-linear effects using the following scheme:
\begin{equation}\label{eq:cl_dec}
    C_{ij,\ell}^{\mathrm{tot}} = C_{ij,\ell}^{\mathrm{lin}} + C_{ij,\ell}^{\delta, \mathrm{limb}} - C_{ij,\ell}^{\mathrm{lin}, \mathrm{limb}}\;.
\end{equation}
This approach ensures high computational efficiency by restricting the non-Limber integration to the linear regime, while still capturing the necessary non-linear physics through the Limber approximation.

We note that because the contributions from PNG and RSD decay rapidly at high multipoles, as will be shown in Sec.~\ref{sec:kernels}, we choose to exclude them from the Limber evaluation for $\ell \geq 200$. While it is possible to implement a Limber-based treatment for these effects \citep{tanidis2019developing}, we find that their exclusion at high $\ell$ significantly lightens the computational load without compromising the required numerical accuracy, as their contribution on the computed statistics is significantly smaller than $1\%$ (for reference, see Fig.~\ref{fig:cl_gg_pieces}).

\section{Extension of the \blast{} algorithm}\label{sec:extensions}
We now generalize the angular power spectrum integral defined in Eq.~\ref{eqn:cls-general} to support a joint analysis across multiple additional observables. To maintain the efficiency of the \blast{} algorithm, we must ensure that new scale-dependent terms and derivatives of the Bessel functions are correctly mapped onto the pre-computed basis integrals $\tilde{T}_{n,\ell}$. In this section, we describe the reformulated treatment of the power spectrum and the projection kernels for the new physical probes.

\subsection{Generalized Power Spectrum Treatment}
A key difference with respect to the original \blast{} formulation concerns the treatment of the unequal-time correlator and the handling of the matter power spectrum. In the original \blast{} algorithm, starting from the matter power spectrum evaluated on a grid of wavenumbers and redshifts, $P(k,z)$, the unequal-time galaxy power spectrum was constructed as a geometric mean,
\begin{equation}\label{eq:uneq_old}
    P_{\mathrm{gg}}(k,z_1,z_2)=\sqrt{P_{\mathrm{gg}}(k,z_1)\,P_{\mathrm{gg}}(k,z_2)}\;.
\end{equation}
This approximation breaks down in the presence of PNG. In particular, so-called local-type PNG induces a scale-dependent correction to the scale-independent linear galaxy bias, \(b_1(z)\), parametrized by the local $\fnl$ parameter as
\begin{equation}\label{eqn:bias_total}
    b(k,z)=b_1(z)+2\,\delta_{\rm c}\,\frac{b_1(z)-p}{T_{\Phi\rightarrow\delta}(k,z)}\,\fnl\;,
\end{equation}
where \(b_1\) is the scale-independent linear galaxy bias, \(\delta_{\rm c}\simeq1.69\) is the critical density contrast for collapse, \(p\) will depend in general on the specific tracer under investigation, and \(T_{\Phi\rightarrow\delta}\) is a transfer function (more on this later). Throughout this work, whenever PNG is included, we adopt \(p=1\), corresponding to the standard universality relation $b_\Phi = 2\delta_{\rm c}(b_1-1)$ \citep{slosar2008constraints}.
Since this correction can cause the bias to change sign as a function of scale, the factorized prescription of Eq.~\ref{eq:uneq_old} is no longer generally applicable.

To address this, we reformulate the power spectrum in terms of transfer functions. We define the primordial potential power spectrum as
\begin{equation}
    P_\Phi(k)=\frac{9}{25}\,\frac{2\,\pi^2}{k^3}\,A_{\mathrm{s}}\,\left(\frac{k}{k_{\mathrm{pivot}}}\right)^{n_{\mathrm{s}}-1}\;,
\end{equation}
where \(A_{\rm s}\) is the amplitude of primordial scalar perturbations, \(n_{\rm s}\) their spectral index, and \(k_{\rm pivot}\) a reference scale.
The total matter transfer function is obtained by interfacing \blast{} with an external code to compute $P_{\mathrm{lin}}(k,z)$, and inverting 
the standard relation $P_{\mathrm{lin}}(k,z) = T_{\Phi\rightarrow\delta}^2(k,z)\,P_\Phi(k)$:
\begin{equation}
    T_{\Phi\rightarrow\delta}(k,z) = \sqrt{\frac{P_{\mathrm{lin}}(k,z)}{P_\Phi(k)}}\;.
\end{equation}
In the current implementation, the linear matter power spectrum $P_{\mathrm{lin}}(k,z)$ is obtained through an interface with the emulator \texttt{Mapse.jl}\footnote{\url{https://github.com/CosmologicalEmulators/Mapse.jl}} \citep{bonici2026inprep}, which provides the linear spectrum and a native \julia{} implementation of \texttt{halofit} \citep{takahashi2012revising}. Alternatively, one could interface \blast{} with traditional Boltzmann solvers like \texttt{CLASS} \citep{lesgourgues2011cosmic}, for which a \julia{} wrapper is available\footnote{\url{https://github.com/hersle/CLASS.jl}}, or the \julia{} native Boltzmann solver \texttt{SymBoltz.jl}\footnote{\url{https://github.com/hersle/SymBoltz.jl}} \citep{sletmoen2026symboltz}. Using the transfer-function formulation, the unequal-time matter power spectrum reads
\begin{equation}    P(k,z_1,z_2)=P_\Phi(k)\,T_{\Phi\rightarrow\delta}(k,z_1)\,T_{\Phi\rightarrow\delta}(k,z_2)\;,
\end{equation}
or, equivalently, in the $(\chi,R)$ coordinate system,
\begin{equation}    P(k,\chi,R\chi)=P_\Phi(k)\,T_{\Phi\rightarrow\delta}(k,\chi)\,T_{\Phi\rightarrow\delta}(k,R\chi)\;.
\end{equation}
The transfer functions are interpolated onto the $(\chi,R)$ grid using Akima splines \citep{akima1970new} while the power spectrum is evaluated on a $k$-grid of Chebyshev points to perform the FFT-based extraction of the coefficients $c_n^{\mathrm{AB}}$.

\subsection{Numerical Infrastructure: Building blocks}
The expansion to a larger set of observables introduces significant numerical complexity relative to the original implementation, where the algorithm was dealing with one set of Chebyshev coefficients and three precomputed $\tilde T$ functions, for a total of $3$ projected matter densities $w_\ell(\chi, R\chi)$. The complexity arises from two primary sources: the scale dependent nature of the PNG contribution and the second derivative of the Bessel functions introduced by RSD, as we shall shortly see.

First, since the scale dependent bias scales as $1/T_{\Phi\rightarrow\delta}(k,z)$, PNG introduces contributions that cannot be described by a single unequal-time matter power spectrum. This requires \blast{} to handle three distinct classes of spectra: the full unequal-time spectrum (two transfer functions), mixed spectra (single transfer function), and the primordial potential spectrum $P_\Phi(k)$ itself.
Each power spectrum must be decomposed independently in $k$-space and treated as a separate building block in the computation, leading to three sets of Chebyshev coefficients.

Secondly, RSD contributions involve second derivatives of spherical Bessel functions with respect to their argument, $\de^2j_\ell/\de x^2\equiv j_\ell''$, which we implement using the recursion relation
\begin{equation}\label{eqn:bessel_der}
    j_\ell''(x)=\frac{2}{x}j_{\ell+1}(x)+\frac{\ell^2-\ell-x^2}{x^2}j_\ell(x)\;.
\end{equation} 
This requires the precomputed $\tilde{T}$-functions to account for multiple combinations of Bessel derivative indices $(\alpha_1,\alpha_2) = \{0, 2\}$ and varying $k$-scalings. In the original formulation, only the $\alpha_1=\alpha_2=0$ case was required, involving $3$ precomputed $\tilde{T}$'s (see Eq.~33 of \citep{chiarenza2024blast}). The extended version increases this to $8$ distinct $\tilde{T}$'s.

Notably, in the $(\chi, R)$ coordinate system, the $\tilde{T}$-functions are generally not symmetric when $\alpha_1 \neq \alpha_2$, i.e.
\begin{equation}
    \tilde T_\ell^{\alpha_1\alpha_2}(k; \chi,R\chi) \neq \tilde T_\ell^{\alpha_2\alpha_1}(k; R\chi,\chi)\;.
\end{equation}
Consequently, both must be retained to evaluate the integral of Eq.~\ref{eqn:final_cls}, which is better expressed as
\begin{multline}
    C_{ij,\ell}^{\mathrm{AB}} = \int_0^\infty \de \chi\, \chi \int_0^1 \de R\, 
    \Big[ \mathcal{K}_i^{\mathrm{A}}(\chi)\,\mathcal{K}_j^{\mathrm{B}}(R\chi)\, w_{\mathrm{AB},\ell}^{\alpha_1\alpha_2}(\chi,R\chi) \\
    + \mathcal{K}_j^{\mathrm{B}}(\chi)\,\mathcal{K}_i^{\mathrm{A}}(R\chi)\, w_{\mathrm{AB},\ell}^{\alpha_2\alpha_1}(R\chi,\chi) \Big]\;.
\end{multline}
Taken together, these generalizations lead to a total of $17$ projected matter densities $w_{AB, \ell}^{\alpha_1\alpha_2}(\chi, R\chi)$. These constitute the numerical building blocks from which all auto- and cross-correlations are assembled. A complete summary of these building blocks is provided in Appendix~\ref{app:contributions}.

\subsection{Physical Observables and Kernels}\label{sec:kernels}
In this section, we define the specific projection kernels for the observables supported by \blast{}. These kernels are mapped onto the $17$ projected matter densities described in the previous subsection to construct the final angular power spectra.

\subsubsection{Galaxy clustering}
The galaxy clustering observable accounts for number counts, redshift-space distortions, magnification bias, and primordial non-Gaussianity, with the following kernels
\begin{align}\label{eq:kernels}
    W_i^{\delta}(z)&=\frac{H(z)}{c}\,b_{1}(z)\,n_i(z)\;,\\
    W_i^{\mathrm{RSD}}(z)&=\frac{H(z)}{c}\,f(z)\,n_i(z)\;,\\
    W_i^{\mu}(z) &= \frac{3}{2}\,\frac{H_0^2\,\Omega_{\rm m}}{c^2}\,\chi(z)\,(1+z)\nonumber\\ 
    &\quad\times\int_z^\infty \de z'\, n_i(z')\,\frac{\chi(z')-\chi(z)}{\chi(z')}\,2\left[\mathcal{Q}(z)-1\right]\;,\label{eq:kernel_mag}\\
    W_i^{\mathrm{PNG}}(z)&=\frac{H(z)}{c}\,\fnl\,
    2\,\delta_{\rm c}\,[b_1(z)-p]
    \,n_i(z)\;,\label{eq:kernel_PNG}
\end{align}
where \(f(z)\equiv-\de\ln D(z)/\de\ln(1+z)\) is the growth rate, given \(D(z)\) the growth factor, and \(\mathcal{Q}(z)\) is the magnification bias parameter, defined in terms of the physical number of sources per unit redshift per solid angle, \(n(z;\le m_{\rm c})\), and the magnitude limit of the survey, \(m_{\rm c}\), as \(\mathcal{Q}=\frac{5}{2}\pd\log_{10}n/\pd m_{\rm c}\) \citep[see e.g.][]{maartens2021magnification, scranton2005detection, mahony2022forecasting,wenzl2024}.

The magnification bias kernel of Eq.~\ref{eq:kernel_mag}, the weak lensing kernel of Sec.~\ref{sec:kernels} (which adopts the same form with $\mathcal{Q}(z)\equiv 3/5$), and the ISW kernel of Eq.~\ref{eqn:isw_kernel} are written in terms of $H_0^2$ and $\Omega_{\rm m}$, assuming a standard matter background scaling as $\bar\rho_{\rm m}(z) = \rho_{\rm m,0}(1+z)^3$. More generally, this combination can be replaced via
\begin{equation}
    \frac{3}{2}\frac{H_0^2\Omega_{\rm m}}{c^2}(1+z) \to 4\pi G\,\bar\rho_{\rm m}(z)\,(1+z)^{-2}\;,
\end{equation}
which allows these kernels to be applied to scenarios with non-standard matter background evolution. The current \blast{} implementation assumes the standard $\Lambda$CDM scaling.


As the definition of a PNG kernel is not standard, in the following we motivate our formalism. The impact of PNG is typically implemented as a scale-dependent correction to the linear galaxy bias, as per Eq.~\ref{eqn:bias_total}. Starting from the general expression for the angular power spectrum in Eq.~\ref{eqn:cls-general}, we note that PNG affects only the galaxy number counts, entering through this scale-dependent correction. For the galaxy number counts auto-correlation, Eq.~\ref{eqn:cls-general} becomes:
\begin{multline}
    C_{ij,\ell}^{\delta\delta} =
    \int_0^\infty \de \chi_1 \, W_i^{\delta}(\chi_1) \,
    \int_0^\infty \de \chi_2 \, W_j^{\delta}(\chi_2) \\
    \times \int_0^\infty \de k \, k^2\,
    P_{\mathrm{mm}}(k,\chi_1,\chi_2)\,
    j_\ell(k\chi_1)\,j_\ell(k\chi_2)\;,
\end{multline}
where $W_i^\delta(\chi) = (H(z)/c)\,b_1(z)\,n_i(z)$ as defined in Eq.~\ref{eq:kernels}. In the presence of PNG, the scale-independent bias $b_1$ receives a scale-dependent correction. More explicitly, we can write:
\begin{multline}
    C_{ij,\ell}^{(\delta+\mathrm{PNG})(\delta+\mathrm{PNG})} =
    \int_0^\infty \de \chi_1 \, \frac{H(\chi_1)}{c}\,n_i(\chi_1) \\
    \times \int_0^\infty \de \chi_2 \, \frac{H(\chi_2)}{c}\,n_j(\chi_2) \\
    \times \int_0^\infty \de k \, k^2\,
    P_{\mathrm{gg}}(k,\chi_1,\chi_2)\,
    j_\ell(k\chi_1)\,j_\ell(k\chi_2)\;,
\end{multline}
where $P_{\mathrm{gg}}$ now encodes the full scale-dependent bias via Eq.~\ref{eqn:bias_total}.

Because \blast{} performs the $k$-space decomposition prior to the line-of-sight integration, PNG effects must be incorporated at the level of the projected matter densities, allowing the $\fnl$ dependence to be enclosed into a dedicated kernel.
To see how this is achieved, we expand the galaxy auto-spectrum and change to the $(\chi, R)$ coordinates (where $\chi_2 = R\chi_1$, as introduced in Eq.~\ref{eqn:final_cls}):
\begin{align}
    P_{\mathrm{gg}}(k,\chi,R\chi)
    &=
    b_1^2\,P_\Phi(k)\,T_{\Phi\rightarrow\delta}(k,\chi)\,
    T_{\Phi\rightarrow\delta}(k,R\chi)\nonumber \\
    &\quad
    + \,b_1\,\fnl\,2\delta_{\rm c}(b_1-p)\,
    P_\Phi(k)\,T_{\Phi\rightarrow\delta}(k,R\chi)\nonumber \\
    &\quad
    + \,b_1\,\fnl\,2\delta_{\rm c}(b_1-p)\,
    P_\Phi(k)\,T_{\Phi\rightarrow\delta}(k,\chi)\nonumber \\
    &\quad
    + \fnl^2\,[2\delta_{\rm c}(b_1-p)]^2\,P_\Phi(k).\;
\end{align} 
The four terms in this equation clearly show how three distinct classes of power spectra must be decomposed independently in $k$-space: the full unequal-time matter spectrum involving two transfer functions, 
mixed spectra involving a single transfer function, and the primordial spectrum $P_\Phi(k)$ alone. Crucially, the mixed terms are not symmetric under exchange of $\chi$ and $R\chi$, so the cross-contributions $C_\ell^{\delta-\mathrm{PNG}}$ and 
$C_\ell^{\mathrm{PNG}-\delta}$ must be treated as distinct building blocks. In practice, this does not require additional FFTs, but it does increase the total number of required $w_\ell$ (see Appendix~\ref{app:contributions} for more details).
Defining $W_i^{\mathrm{PNG}}$ as in Eq.~\ref{eq:kernel_PNG}, the number counts angular power spectrum, in the presence of PNG, decomposes as
\begin{equation}
    C_\ell^{(\delta+\mathrm{PNG})(\delta+\mathrm{PNG})} = 
    C_\ell^{\delta\delta} + C_\ell^{\delta-\mathrm{PNG}} + 
    C_\ell^{\mathrm{PNG}-\delta} + C_\ell^{\mathrm{PNG}-\mathrm{PNG}}\;,
\end{equation}
where each term is computed as an independent building block within \blast{}. This reformulation transforms the scale-dependent problem into a sum of standard line-of-sight integrals, allowing PNG to be evaluated with the same efficiency as the density or RSD contributions, 
while correctly accounting for the non-trivial $(\chi, R)$ symmetries of the mixed terms. The interaction of the PNG kernel with RSD, magnification bias and the other probes is detailed in Appendix~\ref{app:contributions}.

Fig.~\ref{fig:cl_gg_pieces} illustrates the fractional contribution of each component to the galaxy clustering angular power spectrum for a single Gaussian redshift bin centered at $\bar{z}=1.5$ with $\sigma_z=0.15$. While the density term $\delta$ dominates across most scales, RSD and magnification bias provide significant corrections at large angular scales. Notably, for $\fnl=20$, the PNG contribution accounts for nearly $50\%$ of the total power at $\ell=2$, underscoring the necessity of an accurate beyond-Limber treatment at the largest scales probed by Stage-IV surveys.
\begin{figure}[t]
    \centering
    \includegraphics[width=\columnwidth]{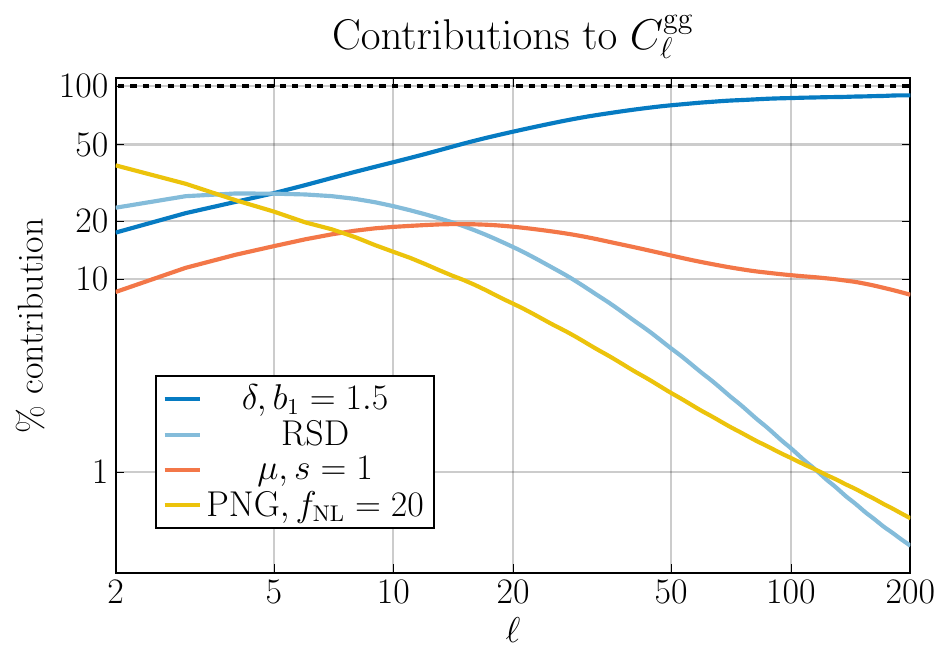}
    \caption{Relative physical contributions to the galaxy clustering angular power spectrum $C_\ell^{\mathrm{gg}}$ for a Gaussian redshift bin centered at $\bar{z}=1.5$ with $\sigma_z=0.15$. Each curve shows the fractional contribution of a single physical component to the total signal: number counts ($\delta$, with linear bias $b_1=1.5$), redshift-space distortions (RSD), magnification bias ($\mu$, with slope parameter $s=1$, related to the magnification bias parameter via $\mathcal{Q} = 5s/2$), and primordial non-Gaussianity (PNG, with $\fnl=20$). The black dashed line shows the sum of all contributions. At the largest scales ($\ell \approx 2$), PNG and RSD provide substantial contributions, making a beyond-Limber treatment essential for accurate modeling.}
    \label{fig:cl_gg_pieces}
\end{figure}

\subsubsection{Weak lensing}
\begin{figure}[h!]
    \centering
    \includegraphics[width=\columnwidth]{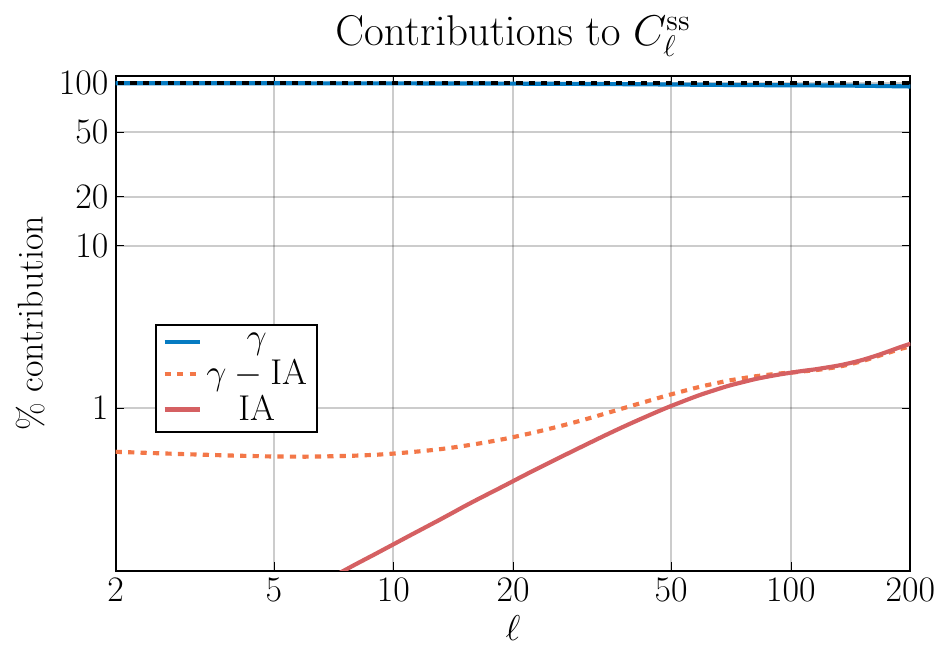}
    \caption{Relative physical contributions to the cosmic shear angular power spectrum $C_\ell^{\mathrm{ss}}$ for a Gaussian redshift bin centered at $\bar{z}=1.5$ with $\sigma_z=0.15$. Each curve shows the fractional contribution of a single physical component to the total signal: cosmic shear ($\gamma$), intrinsic alignments (using the NLA model), and their combined cross-correlation ($\gamma-\text{IA} + \text{IA}-\gamma$). The black dashed line shows the sum of all contributions. The cross-correlation term is negative and is shown as a dashed colored line, with its absolute value plotted. The shear signal dominates across all scales, with IA contributing at the $\approx 1\%$ level around $\ell = 200$.}
    \label{fig:cl_ss_pieces}
\end{figure}
In the original \blast{} implementation, the weak lensing observable included only cosmic shear, described by the kernel \(W_i^{\rm s}\), defined as in Eq.~\ref{eq:kernel_mag} with \(\mathcal{Q}(z)\equiv3/5\). 
The extended version introduced here incorporates intrinsic alignments (IA) \citep{hirata2004intrinsic, hirata2004galaxy} in a deliberately general form. 

If the IA amplitude is not specified, \blast{} uses the non-linear alignment (NLA) model \citep{catelan2001intrinsic, hirata2007intrinsic} as the default prescription. NLA is one of the standard IA models adopted as default in large weak lensing analyses \citep[e.g.][]{abbott2022dark, 
abbott2026dark, wright2025kids, dalal2023hyper}, alongside 
the more general TATT model \citep{blazek2019beyond, schmitz2018time}. In this case, the redshift-dependent IA amplitude is automatically computed as
\begin{equation}
    A_{\mathrm{IA}}(z) = -A\,\mathcal{C}\,\frac{\Omega_{\rm m}}{D(z)}\;,
\end{equation}
where $D(z)$ denotes the linear growth factor, $\mathcal{C} \equiv C_1\rho_{\rm crit} = 0.0134$ \citep{bridle2007dark} with $C_1 = 5\times10^{-14}\,M_\odot^{-1} 
h^{-2}\,\mathrm{Mpc}^3$, and $A = 1.72$ , following Eq.~6 of \citep{joachimi2011constraints} and matching the implementation in \texttt{CCL} \citep{chisari2019core}.

Alternatively, the algorithm can take as input a custom redshift-dependent amplitude, $A_{\mathrm{IA}}(z)$. In either case, the IA kernel is constructed as
\begin{equation}
    W_i^{\mathrm{IA}}(z) = \frac{H(z)}{c}\,A_{\mathrm{IA}}(z)\,n_i(z)\;.
\end{equation}
This design enables the straightforward implementation of a wide range of intrinsic alignment models without requiring any modification to the core integration machinery.

Fig.~\ref{fig:cl_ss_pieces} shows the fractional contributions to the cosmic shear auto-correlation $C_\ell^{\mathrm{ss}}$. The signal is dominated by the shear kernel $\gamma$ across all scales, with IA contributing at the $\approx 1\%$ level around $\ell=200$.
\newpage
\subsubsection{CMB probes}
\begin{figure}[h!]
    \centering
    \begin{minipage}[b]{\columnwidth}
        \centering
        \includegraphics[width=\linewidth]{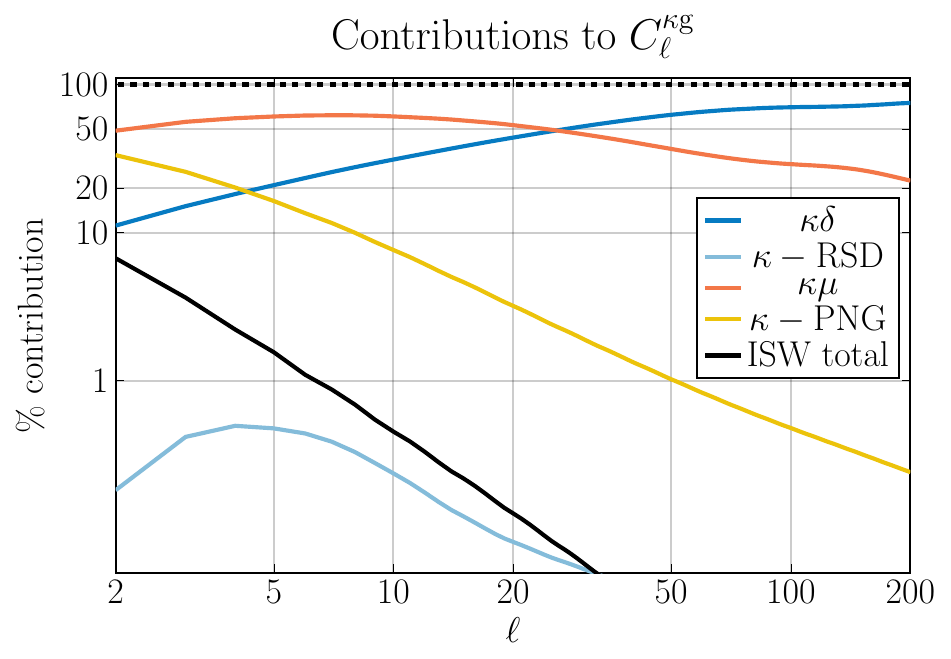}
        \caption{Relative physical contributions to the galaxy-CMB lensing cross-correlation $C_\ell^{\kappa\mathrm{g}}$, decomposed into number counts ($\kappa\delta$), RSD ($\kappa-\mathrm{RSD}$), magnification bias ($\kappa\mu$), PNG ($\kappa-\mathrm{PNG}$, with $\fnl=20$), and the ISW total contribution. The black dashed line shows the sum of all contributions. Dashed colored lines indicate negative contributions, shown as absolute values.}
        \label{fig:cl_kg_pieces}
    \end{minipage}
    \hfill
    \begin{minipage}[b]{\columnwidth}
        \centering
        \includegraphics[width=\linewidth]{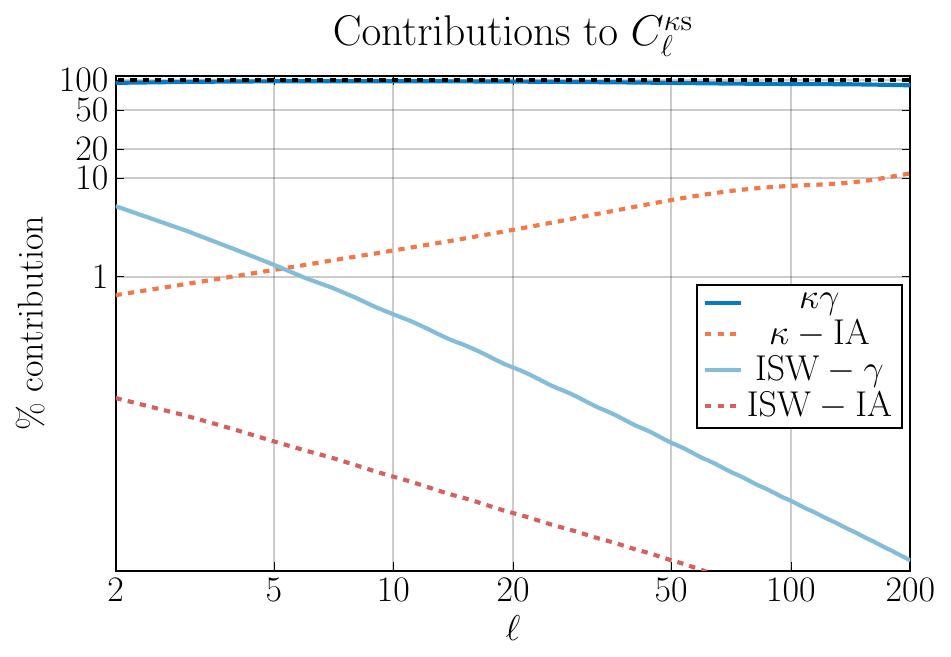}
        \caption{Relative physical contributions to the weak lensing-CMB cross-correlation $C_\ell^{\kappa\mathrm{s}}$, decomposed into shear ($\kappa\gamma$), IA ($\kappa-\mathrm{IA}$), and their respective ISW cross-terms (ISW$-\gamma$, ISW$-$IA). The black dashed line shows the sum of all contributions. Dashed colored lines indicate negative contributions, shown as absolute values.}
        \label{fig:cl_ks_pieces}
    \end{minipage}
\end{figure}
The extended version of \blast{} includes CMB lensing as an additional cosmological probe. The dominant contribution is the CMB lensing convergence, whose kernel is given by
\begin{equation}
    W^{\kappa}(z)
    =
    \frac{3}{2}\,\frac{H_0^2\,\Omega_{\rm m}}{c^2}\,
    \chi(z)\,(1+z)\,\left[1-\frac{\chi(z)}{\chi_{\mathrm{CMB}}}\right]\;,
\end{equation}
where $\chi_{\mathrm{CMB}}=\chi(z=1090)$ denotes the comoving distance to the surface of last scattering.

In addition, \blast{} supports the ISW effect \citep{isw}, implemented through the kernel
\begin{equation}\label{eqn:isw_kernel}
    W^{\mathrm{ISW}}(z)
    =
    3\,T_{\mathrm{CMB}}\,
    \frac{H_0^2\,\Omega_{\rm m}}{c^3}\,
    H(z)\,[1-f(z)]\;,
\end{equation}
where we fix $T_{\mathrm{CMB}}=2.7255\,\mathrm{K}$ \citep{fixsen2009temperature}. The ISW effect probes the time evolution of gravitational potentials on large scales and is therefore particularly sensitive to 
the low-$\ell$ regime (see Fig.~\ref{fig:cl_kg_pieces}).

The current implementation does not compute the CMB lensing auto-spectrum $C_\ell^{\kappa\kappa}$. Accurately evaluating this quantity would require extending the radial integration grid well beyond $z = 3.5$, well beyond the redshift range relevant for the other probes, resulting in a substantial loss of computational efficiency. Since $C_\ell^{\kappa\kappa}$ can be obtained accurately using fast emulators or Boltzmann solvers, it is excluded from the 
present implementation.

Fig.~\ref{fig:cl_kg_pieces} and \ref{fig:cl_ks_pieces} show the fractional contributions to the galaxy-CMB lensing cross-correlation $C_\ell^{\kappa\mathrm{g}}$ and the weak lensing-CMB cross-correlation $C_\ell^{\kappa\mathrm{s}}$, respectively. In both cases, the ISW effect and RSD are prominent at low multipoles: for $C_\ell^{\kappa\mathrm{g}}$, the ISW contribution is particularly significant at $\ell \lesssim 10$, while for $C_\ell^{\kappa\mathrm{s}}$, it reaches $\approx 10\%$ of the total power at the largest scales. Intrinsic alignments have a more significant influence at higher multipoles in the cross-correlation, reaching 
$\approx 10\%$ at $\ell=200$, compared to $\approx 1\%$ for the shear auto-correlation. 

\section{Performance and Validation}\label{sec:performance}
To establish a baseline for the extended version of \blast{}, we first perform a regression test against the \nk{} challenge requirements \citep{leonard2022n5k}. Throughout this paper, we assume a standard $\Lambda$CDM cosmology, as in \citep{leonard2022n5k},
\begin{align*}
&\{h,\,\Omega_{\rm b},\,\Omega_{\rm c},\,10^9\,A_{\rm s},\,n_{\rm s}\} \\
&=\{0.6727,\,0.0492,\,0.2649,\,2.121,\,0.9645\}.
\end{align*}
We verify that the accuracy requirement of $\Delta\chi^2 < 0.2$ in the range $2 < \ell < 200$ for an ideal LSST 10-year scenario (Y10) is still satisfied: using the same hyperparameters as in \citep{chiarenza2024blast}, the updated implementation achieves $\Delta\chi^2 = 0.195$. This benchmark focuses on the non-Limber regime where beyond-Limber accuracy is most critical, however \blast{} works up to $\ell \leq 2000$ as shown in the likelihood analysis of Sec.~\ref{sec:likelihood}. 
Since the code was completely restructured and now handles more complex operations, we also timed the algorithm in this N5K challenge configuration; the updated results are presented in Table~\ref{tab:fid_res}. The \blast{} timings are higher than those reported in Table~1 of \citep{chiarenza2024blast}, reflecting the additional bookkeeping overhead introduced by the extended framework even when running in the simplified N5K configuration.  Compared to Table~1 of \citep{chiarenza2024blast}, we include in the comparison the recently introduced \swift{} code \citep{Reymond_2026}. The \swift{} CPU timings reported in \citep{Reymond_2026} were obtained using a single thread, whereas our timings make full use of $64$ available CPU threads. To place the two codes on a comparable footing, we rescale the \swift{} timings to $64$ threads by applying the average performance ratio between our hardware and theirs for the \texttt{FKEM}, \texttt{Levin}, and \texttt{matter} entries, yielding a consistent $5\times$ speedup factor. This procedure assumes an ideal linear scaling with thread count. We note that this assumption of perfect linear scaling is optimistic, so the rescaled \swift{} timings reported here should be regarded as a lower bound on its true $64$-thread wall-clock cost. As shown in the table, \blast{} remains the most efficient non-Limber method overall, demonstrating that the additional observables and effects introduced in this work do not compromise performance.
\begin{table}[h!]
    \centering
    \renewcommand{\arraystretch}{1.3}
    \begin{tabular}{ccc}
         \hline
         Entry name & Threads & Runtime \\
         \hline
         \texttt{FKEM} & 64 & 0.05871 $\pm$ 0.00005 s \\
         \texttt{matter} & 64 & 0.5510 $\pm$ 0.0009 s \\
         \texttt{Levin} & 64 & 5.6 $\pm$ 0.1 s \\
         \texttt{SwiftC}$_\ell$ & 64  & 0.1146 $\pm$ 0.0005 s \\
         \texttt{SwiftC}$_\ell$ & GPU  & 0.006 $\pm$ 0.001 s \\
         \blast{} & 64 & 0.0864 $\pm$ 0.0261 s \\
         \blast{} & 32 & 0.0437 $\pm$ 0.0067 s \\
         \hline
    \end{tabular}
    \caption{Comparison of runtimes for different entries and number of threads. The fiducial results for the \nk{} challenge are with $64$ cores, however \blast{} achieves its best performance with $32$ threads. }
    \label{tab:fid_res}
\end{table}

\subsection{Comparison to Brute-Force Calculations}
To evaluate the absolute accuracy of the extended framework, we compare \blast{} against a high-precision brute-force numerical reference. This reference is computed by performing a direct numerical integration of Eq.~\ref{eqn:cls-general} without the use of the Chebyshev power spectrum decomposition. To ensure absolute convergence of the oscillatory $k$-integral, we utilize a Clenshaw-Curtis quadrature scheme with $N = 2^{16} + 1$ points. The integration is performed over an extended range from $k_{\min} = 5 \times 10^{-5}\,h\,\text{Mpc}^{-1}$ to $k_{\max} = 16\,h\,\text{Mpc}^{-1}$, representing a significant expansion compared to the limits adopted in the original \nk{} challenge \citep{leonard2022n5k} ($k_{\min} = 3.57 \times 10^{-4}\,h\,\text{Mpc}^{-1}$ and $k_{\max} = 15.38\,h\,\text{Mpc}^{-1}$). The lower $k_{\min}$ is specifically chosen to capture the large-scale signatures of primordial non-Gaussianity, ensuring that these contributions are fully resolved within the non-Limber regime. The same extended $k$ range is employed in the computation of the $\tilde T$ functions. This extended range requires a corresponding increase in the number of points used in the Chebyshev approximation of the power spectrum, which is now $n_\mathrm{cheb}=161$ (compared to $n_\mathrm{cheb}=120$, which was sufficient for the \nk{} challenge requirements).

\begin{figure}[t!]
    \centering
    \includegraphics[width=\columnwidth]{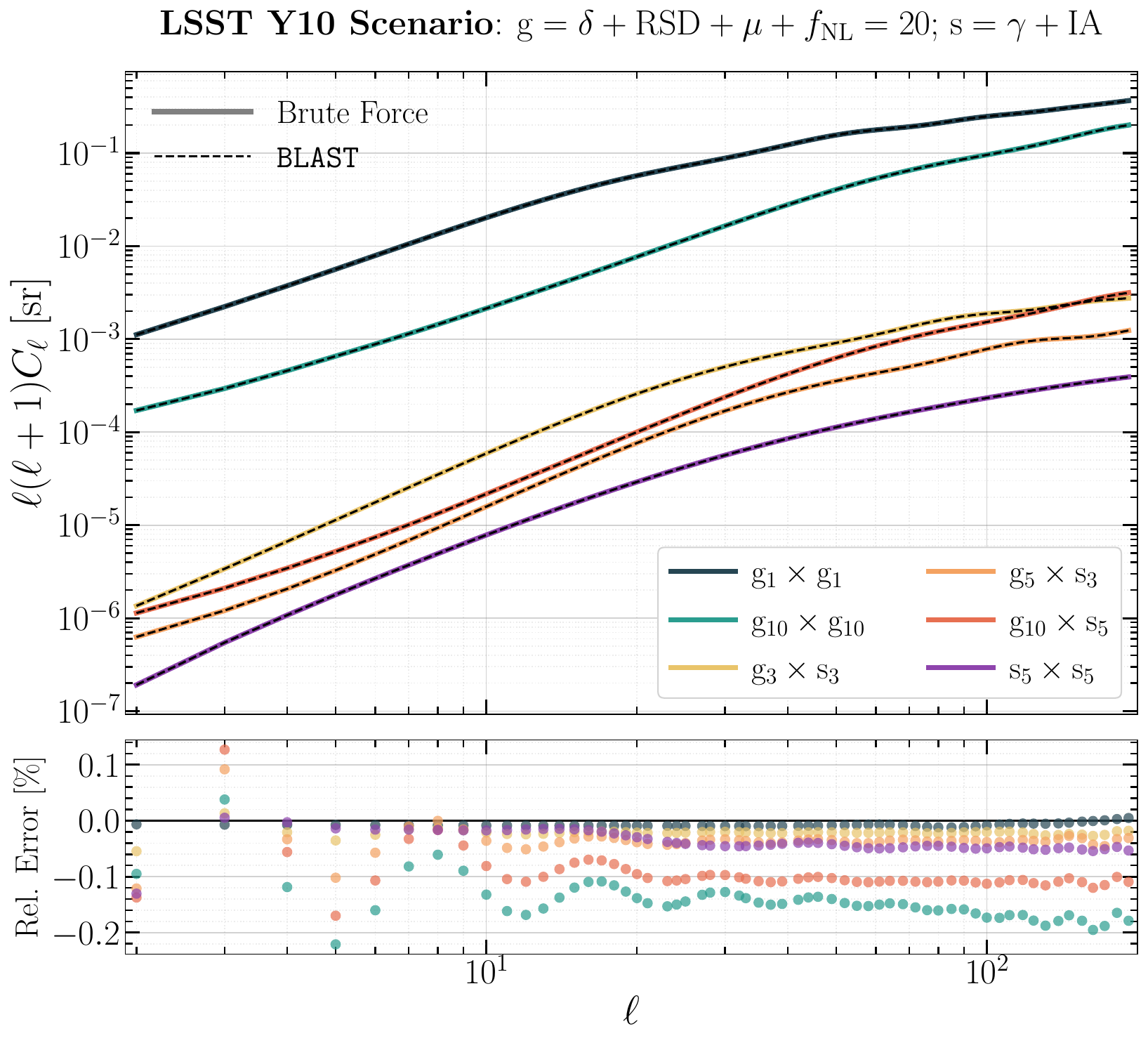}
    \caption{Comparison of angular power spectra between \blast{} and a high-precision brute-force integration for an LSST Y10 scenario ($10$ clustering and $5$ shear bins). The calculation accounts for the full set of physical effects, including number counts, RSD, magnification bias, primordial non-Gaussianity ($\fnl=20$), and intrinsic alignments (IA). The top panel shows the angular power spectra for some of the auto- and cross-correlations. The bottom panel displays the relative error in percentage, demonstrating that \blast{} maintains a sub-percent precision across the non-Limber regime. }
    \label{fig:bf}
\end{figure}

The brute-force calculation makes use of a significantly denser integration grid than the standard \blast{} configuration, with $768$ points in $\chi$ and $384$ points in $R$. The final angular power spectra are obtained by integrating the resulting projected matter densities using Simpson's rule in $\chi$ and a truncated Clenshaw-Curtis scheme in $R$ \citep{chiarenza2024blast}. 

For this validation, we adopt an LSST Y10 configuration following the \nk{} challenge specifications \citep{leonard2022n5k}, consisting of $10$ galaxy clustering and $5$ cosmic shear bins. But now the analysis incorporates the full set of physical effects: RSD, magnification bias, local-type PNG, and IAs modeled via the standard NLA prescription. As in \citep{leonard2022n5k}, the linear galaxy bias values for the $10$ clustering bins are set to 
\begin{align}\label{eqn:bias_numbers}
b_{1i} &=\{1.3767, 1.4512, 1.5284, 1.6080, 1.6896, \nonumber \\
     &\qquad1.7729, 1.8577, 1.9438, 2.0309, 2.1189\}\;,
\end{align}
while the magnification bias parameters are set to \begin{align}
\mathcal{Q}_i &=\{0.412, 0.624, 0.677, 0.825, 0.97, \nonumber \\
     &\qquad0.74, 0.895, 0.99, 1.08, 1.42\}\;,
\end{align}
following \citep{mahony2022forecasting}.
We perform this comparison for three distinct scenarios of primordial non-Gaussianity, setting $\fnl = \{20, 0, -20\}$, adopting $p=1$
throughout. In the fiducial \blast{} configuration employed in this analysis, consisting of $\mathrm{n}\chi=128$, $\mathrm{n}R=64$, and $n_\mathrm{cheb}=161$ points, we find that the difference between \blast{} and the brute-force reference consistently satisfies the \nk{} challenge criterion \citep{leonard2022n5k}. Specifically, we achieve total $\Delta\chi^2$ values of $0.166$, $0.168$, and $0.171$ for $\fnl=20$, $0$, and $-20$, respectively, in the multipole range $2 < \ell < 200$. The fractional differences between the two methods for the case $f_\mathrm{NL}=20$ are shown in Fig.~\ref{fig:bf}, where a slightly larger residual is visible for higher-redshift bin pairs. While this trend is noticeable, it does not affect our conclusions: the total $\Delta\chi^2$ remains well within the target threshold across all configurations. While we suspect the effect may be related to the approximation introduced in Eq.~\ref{eq:pk_dec} for the non-linear correction, we leave a more detailed investigation to future work. This result demonstrates that the generalized Chebyshev decomposition and the $(\chi, R)$ coordinate transformation maintain high numerical fidelity even in the presence of scale-dependent bias and complex beyond-Limber kernels.
\newpage
\subsection{Comparison to other codes}
\begin{table*}[h!]
    \centering
    \begin{tabular}{l cccc ccc cc}
        \hline
        \hline
        & \multicolumn{4}{c}{Galaxy Clustering} & \multicolumn{2}{c}{Weak Lensing} & \multicolumn{2}{c}{CMB} \\
        \cmidrule(lr){2-5} \cmidrule(lr){6-7} \cmidrule(lr){8-9}
        Code & $\delta$ & RSD & Mag & PNG & Shear & IA & $\kappa$ & ISW \\
        \hline
        \blast{} & \checkmark & \checkmark & \checkmark & \checkmark & \checkmark & \checkmark & \checkmark & \checkmark \\
        \texttt{SwiftC}$_\ell$ & \checkmark & \checkmark & \checkmark & \checkmark & \checkmark & \checkmark & \checkmark & \checkmark \\
        \texttt{CCL (FKEM)} & \checkmark & \checkmark & \checkmark & $\dagger$ & \checkmark & \checkmark & \checkmark & $\dagger$ \\
        \texttt{CAMB} & \checkmark & \checkmark & \checkmark & $\dagger$ & \checkmark & \ding{55} & \checkmark & \checkmark \\
        \texttt{CLASS} & \checkmark & \checkmark & \checkmark & \ding{55} & \checkmark & \ding{55} & \checkmark & \checkmark \\
        \texttt{pylevin} & \checkmark & $\dagger$ & $\dagger$ & $\dagger$ & \checkmark & $\dagger$ & \checkmark & $\dagger$ \\
        \hline
        \hline
    \end{tabular}
    \caption{This table summarizes the physical effects and probes supported by each package within their respective beyond-Limber frameworks. A $\dagger$ indicates that the effect is not available out-of-the-box and requires additional implementation (e.g., manual construction of kernels or combinatorics). While \texttt{CAMB} does not natively support PNG, it permits a general scale-dependent bias $b(k,z)$, which allowed us to implement the PNG-induced scale-dependent bias as in Eq.~\ref{eqn:bias_total}.}
    \label{tab:code_capabilities}
\end{table*}
\begin{figure*}[h!]
    \centering
    \includegraphics[width=0.97\textwidth]{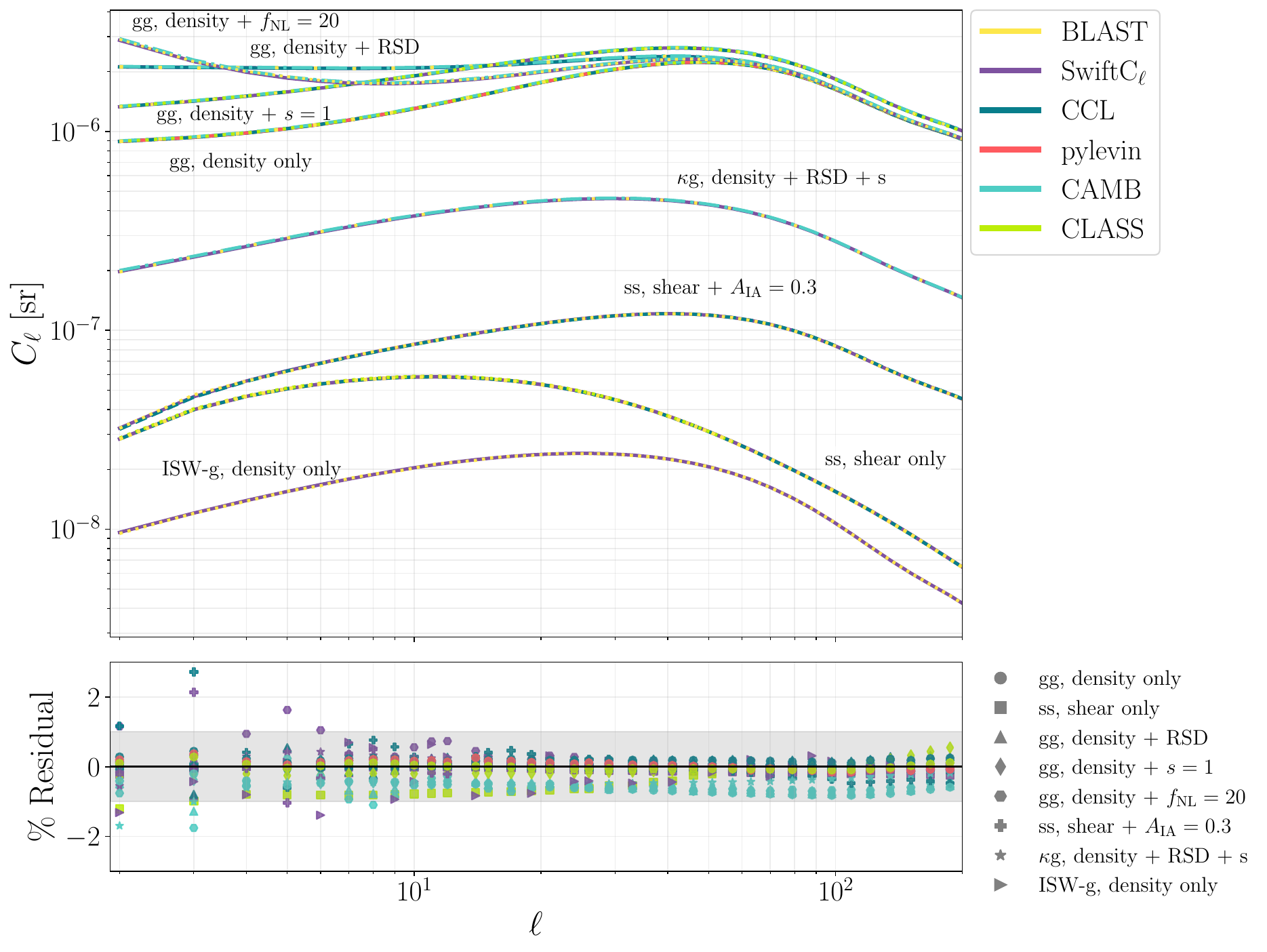}
    \caption{Validation of the \blast{} algorithm against a comprehensive suite of established cosmological codes. We compare angular power spectra across a diverse set of auto- and cross-probes using a Gaussian redshift bin ($\bar{z}=1.5$, $\sigma_z=0.15$) with $b_1=1.5$, $\mathcal{Q}=2.5$, $\fnl= 20$, and $A_{\mathrm{IA}}=0.3$. The top panel illustrates the $C_\ell$ for various physical configurations (see Table.~\ref{tab:code_capabilities} for the specific capabilities of each code). The bottom panel shows the percentage residuals relative to \blast{}, with the gray shaded area denoting a $\pm 1\%$ band. \blast{} demonstrates sub-percent agreement with the consensus of standard libraries across all supported physical effects, confirming its high numerical fidelity even in the highly non-Limber regime where complex kernels like RSD, magnification, and ISW are most prominent. To facilitate a comparison across the full dynamic range, the ISW-$g$ signal is shown as $\ell^{3/2}C_\ell$. An extended version of this plot, showing the residuals in greater detail, is available in Appendix~\ref{app:master_comparison}.}
    \label{fig:master_plot}
\end{figure*}

A central requirement for the extended version of \blast{} is that its new architecture remains fully consistent with established and validated cosmological tools. To demonstrate this, we perform a systematic validation against several independent frameworks summarized in Table~\ref{tab:code_capabilities}, which includes methods that make non-Limber prediction like \texttt{CAMB} \citep{lewis2000efficient} and \texttt{CLASS} \citep{lesgourgues2011cosmic}, as well as specialized beyond-Limber libraries such as \texttt{SwiftC}$_\ell$ \citep{Reymond_2026}, the Core Cosmology Library (\texttt{CCL}) \citep{chisari2019core}, which implements the \texttt{FKEM} algorithm \citep{fang2020beyond}, and \texttt{pylevin} \citep{reischke2025pylevin}. While \texttt{CLASS} and \texttt{CAMB} are not specifically optimized for beyond-Limber performance, we use them as highly reliable benchmarks to verify our implementation.

For this benchmark, we adopt a simplified scenario: a single Gaussian redshift bin centered at $\bar{z}=1.5$ with $\sigma_z=0.15$, assuming $b_1=1.5$, $\mathcal{Q}=2.5$, and $A_{\mathrm{IA}}=0.3$. The benchmark is performed using our baseline cosmology, with the addition of massive neutrinos ($m_\nu=0.06$ eV). All codes rely on the \texttt{CAMB} linear power spectrum with the non-linear part given by the \texttt{halofit} model \citep{takahashi2012revising}, called in \texttt{CAMB} by the `\texttt{takahashi}' keyword, and in \texttt{CLASS} as `\texttt{halofit}'. While \texttt{CLASS} utilizes its own power spectrum, we have carefully matched it to the \texttt{CAMB} baseline. However, subtle differences in the \texttt{halofit} neutrino treatment between the two solvers lead to discrepancies on small scales, which are most evident in the shear-shear auto-correlation; we therefore implement a scale cut for the \texttt{CLASS} comparison, showing auto spectra only for $\ell < 100$.

As illustrated in Fig.~\ref{fig:master_plot}, \blast{} is in excellent agreement with the consensus of established codes. A small number of points exceed the $1\%$ residual threshold, but these are concentrated at the largest scales ($\ell \lesssim 10$) where cosmic variance dominates and percent-level agreement is not required for unbiased parameter inference, and in configurations where the spread among the external codes themselves is comparable to or larger than the residual with respect to \blast{}, indicating that these discrepancies reflect inter-code differences rather than a limitation of \blast{}. An extended version of this plot, showing the residuals in greater detail, is available in Appendix~\ref{app:master_comparison}. We note that while \texttt{CAMB} does not natively support PNG, it permits a general scale-dependent bias $b(k,z)$, which allowed us to implement the PNG-induced scale dependent bias as in Eq.~\ref{eqn:bias_total} for validation purposes.

We further compare against \texttt{pylevin} \citep{reischke2025pylevin}, a general-purpose numerical integrator for multi-Bessel integrals. Designed as a numerical engine rather than a dedicated cosmological library, \texttt{pylevin} does not feature a native interface for projection kernels or power spectra. However, its underlying algorithm is mathematically capable of evaluating any product of Bessel functions, meaning that all physical effects listed in Table~\ref{tab:code_capabilities} are theoretically accessible within this framework, although implementing them within a standard cosmological pipeline requires significant manual handling of kernels and combinatorics. A similar argument applies to the FKEM implementation in \texttt{CCL}: effects such as PNG and ISW are not currently supported as they have not yet been implemented. Scale-dependent contributions such as PNG can be accommodated by subdividing each redshift bin into narrower slices and adopting a scale-independent approximation within each slice, as described in the original method paper \citep{fang2020beyond}, while the ISW effect enters through a standard projection kernel and requires no further approximation. Nevertheless, as the primary goal of our benchmark is to compare the `out-of-the-box' functionality and efficiency of specialized cosmological tools, we restrict our numerical comparison with \texttt{pylevin} and \texttt{CCL} to the effects currently available in each code. For \texttt{pylevin}, the corresponding kernels are interfaced directly from \texttt{CCL}.
It is important to emphasize that while Boltzmann solvers provide a highly reliable benchmark, achieving the necessary non-Limber accuracy requires pushing their accuracy settings to a regime where evaluation times reach $\mathcal{O}(\text{seconds})$. In contrast, \blast{} achieves high precision with the millisecond-level efficiency required for modern, gradient-based inference.

\subsection{Hyperparameters}\label{sec:hyperparameters}
\begin{figure}[h!]
    \centering
    \includegraphics[width=0.97\columnwidth]{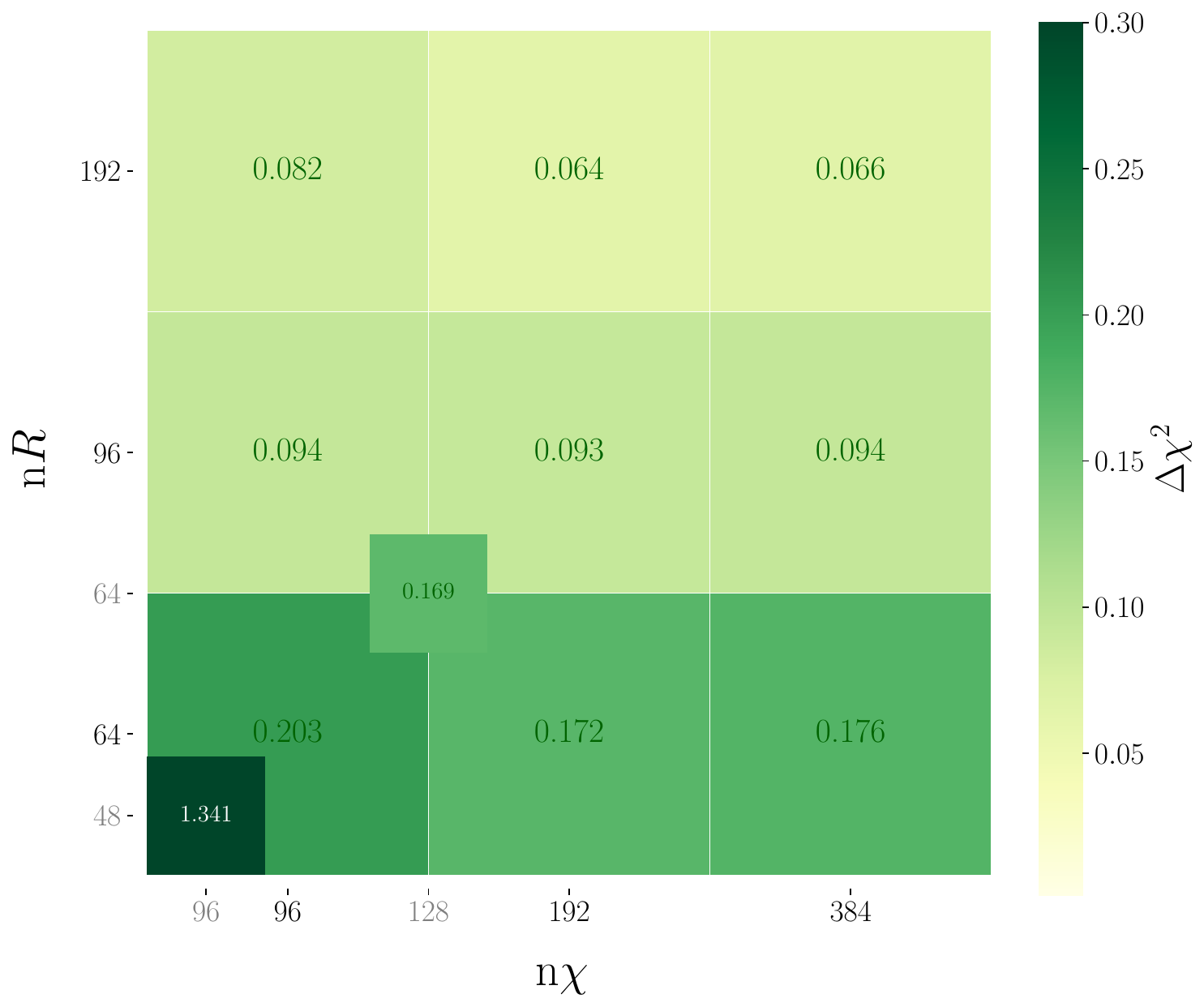}
    \caption{Total $\Delta\chi^2$ between \blast{} and the brute-force reference (Sec.~\ref{sec:performance}) as a function of the hyperparameters $\mathrm{n}\chi$ and $\mathrm{n}R$, evaluated in the multipole range $2 < \ell < 200$ for the full effect scenario (number counts, RSD, magnification bias, cosmic shear, and intrinsic alignments), at fixed $\mathrm{n}_{\mathrm{cheb}} = 161$. The two smallest cells correspond to $(\mathrm{n}\chi, \mathrm{n}R) = (96, 48)$ and $(\mathrm{n}\chi, \mathrm{n}R) = (128, 64)$, which appear smaller due to the non-uniform axis spacing. The coarsest point $(\mathrm{n}\chi, \mathrm{n}R) = (96, 48)$, which was the default configuration in the original \blast{} paper \citep{chiarenza2024blast}, yields $\Delta\chi^2 = 1.341$, now above the target threshold of $\Delta\chi^2 < 0.2$ due to the extra complexity of the setup. Interpolating between the two neighboring grid points to meet this threshold motivates our new default of $(\mathrm{n}\chi, \mathrm{n}R) = (128, 64)$, which achieves $\Delta\chi^2 \approx 0.169$.}
    \label{fig:hyperparam_chi2}
\end{figure}

The accuracy and computational cost of \blast{} are jointly controlled by three hyperparameters: $\mathrm{n}\chi$ and $\mathrm{n}R$, the number of sampling points along the line-of-sight coordinate $\chi$ and the ratio coordinate $R$ (Sec.~\ref{sec:original_blast}), and $\mathrm{n}_{\mathrm{cheb}}$, the order of the Chebyshev expansion of the power spectrum (Eq.~\ref{eq:pk_dec}). To select the optimal operating point for the extended framework, we explore an $11$-point grid in $(\mathrm{n}\chi, \mathrm{n}R)$, fixing $\mathrm{n}_{\mathrm{cheb}} = 161$ and $k_{\min} = 5\times10^{-5}\,h\,\mathrm{Mpc}^{-1}$ to $k_{\max} = 16\,h\,\mathrm{Mpc}^{-1}$, matching the fiducial configuration adopted in Sec.~\ref{sec:performance}. For each grid point, we evaluate the full model, including RSD, magnification bias, and intrinsic alignments alongside number counts and cosmic shear, and compare the resulting angular power spectra against the brute-force reference, computing the total $\Delta\chi^2$ in the multipole range $2 < \ell < 200$. The resulting accuracy map is shown in Fig.~\ref{fig:hyperparam_chi2}: as expected, $\Delta\chi^2$ decreases monotonically as both $\mathrm{n}\chi$ and $\mathrm{n}R$ increase, from $1.341$ at the coarsest grid point to $\approx 0.06$ at the finest tested configurations.

The coarsest grid point $(\mathrm{n}\chi, \mathrm{n}R) = (96, 48)$, which was the default in the original \blast{} paper \citep{chiarenza2024blast}, now yields $\Delta\chi^2 = 1.341$: well above the target threshold of $\Delta\chi^2 < 0.2$. This demonstrates the increased complexity of the full-effect model relative to the original implementation with only number counts and cosmic shear. Interpolating between the two neighboring grid points that straddle the threshold motivates our new default of $(\mathrm{n}\chi, \mathrm{n}R, \mathrm{n}_{\mathrm{cheb}}) = (128, 64, 161)$, which achieves $\Delta\chi^2 \approx 0.169$, safely below the target. The corresponding precomputed $\tilde{T}$ basis integrals are made available to the user upon installation of \blast{}. Users requiring higher accuracy, or wishing to adopt a different $[k_{\min}, k_{\max}]$ or multipole range, can regenerate the $\tilde{T}$ basis integrals using the routines provided in the package.

\section{Simulated Likelihood Analysis}
\label{sec:likelihood}
In order to demonstrate not only the accuracy of \blast{}, but also its speed and readiness for realistic cosmological inference, we perform a simulated likelihood analysis using LSST Y10-like mock data. As in the rest of the paper, the fiducial cosmology is $\{h, \Omega_\mathrm{b}, \Omega_\mathrm{m}, A_\mathrm{s}, n_\mathrm{s}\} = \{0.6727, 0.0492, 0.3156, 2.12107\times10^{-9}, 0.9645\}$ and we use the \blast{} default configuration with $(\mathrm{n}\chi, \mathrm{n}R, \mathrm{n}_{\mathrm{cheb}}) = (128, 64, 161)$. We generate the data following the LSST Dark Energy Science Collaboration Science Requirements Document \citep{mandelbaum2018lsst}. We consider a galaxy clustering (lens) sample, split into $10$ tomographic bins with $n_{\mathrm{gal}} = 40\,\mathrm{arcmin}^{-2}$, and a weak lensing (source) sample, split into $5$ tomographic bins with $n_{\mathrm{gal}} = 27\,\mathrm{arcmin}^{-2}$. Both samples span the redshift range $0 < z < 3.5$. For cosmic shear, we assume an intrinsic ellipticity dispersion of $\sigma_e = 0.28$.

We obtained the synthetic data with \blast{}, using \texttt{Mapse.jl}\footnote{\url{https://github.com/CosmologicalEmulators/Mapse.jl}} for the underlying linear and non-linear matter power spectrum (with the \texttt{halofit} prescription), the same configuration used in the inference. This deliberately controlled setup is intended to assess the numerical and sampling performance of the inference pipeline rather than its robustness to theory-model misspecifications. In particular, because the inference and data-generating models share the same theory pipeline, this exercise does not probe parameter biases induced by residual modelling errors. Linear galaxy bias is included and fixed to the same fiducial per-bin values $b_{1,i}$ as in Eq.~\ref{eqn:bias_numbers}. RSD, magnification bias, and intrinsic alignments are all switched off in the mock data, so that the data-generating model includes number counts and cosmic shear alone. We consider a multipole range $2 \le \ell < 2000$. The power spectra are binned into $24$ bandpowers per probe pair: $8$ linear bins with width $\Delta\ell = 30$ for $0 \le \ell \le 240$, followed by $16$ logarithmic bins for $240 \le \ell < 2000$ with $\Delta\log_{10}(\ell) = 0.055$. Bandpower window functions are top-hat in $\ell$, $\mathcal{B}[i,\ell] = 1/(\ell_{i+1} - \ell_i)$ for $\ell$ within bin $i$ and zero otherwise.

We estimate the covariance matrix under the assumption of Gaussianity, neglecting higher order contributions. These contributions can be relevant for a realistic LSST analysis, and we do not attempt to model or quantify them in the present demonstration. Our aim here is instead to provide a controlled test of the numerical and sampling performance of the \blast{} inference pipeline. Following the philosophy of the \nk{} challenge \citep{leonard2022n5k}, the Gaussian covariance provides a stringent benchmark by yielding tighter statistical uncertainties than a more complete covariance treatment. The posterior widths obtained below should therefore not be interpreted as realistic LSST Y10 forecasts. We adopt the Knox formula \citep{knox1995determination} for Gaussian covariance, which accounts for the observed sky fraction $f_{\rm sky} = 0.4$. Explicitly, the covariance is given by
\begin{equation}
\mathrm{Cov}(C_q^{ab}, C_{q'}^{cd}) =
\frac{\tilde{C}_q^{ac}\tilde{C}_q^{bd} + \tilde{C}_q^{ad}\tilde{C}_q^{bc}}
{f_{\rm sky} N_{\ell\in q}(2q+1)}\,\delta_{qq'} \, ,
\end{equation}
where $q$ is the effective angular mode, $N_{\ell \in q}$ is the number of modes per bin, and $\tilde{C}_q = C_q + N_q$ contains both the binned signal $C_q$ and the noise $N_q$. The galaxy clustering noise is well approximated as Poisson noise, $N_i^g = 1/(n_{\rm gal,cl}\,f_i)$, while the shape noise for cosmic shear is $N_i^s = \sigma_e^2/(n_{\rm gal,sh}\,f_i)$, where $f_i = \int n_i(z)\,\de z \,\big/\, \sum_j \int n_j(z)\,\de z$ is the fraction of the total sample assigned to tomographic bin $i$, obtained by integrating that bin's normalized redshift distribution.

For the inference, we use a subset of the full data vector: the $10$ diagonal auto-correlations of galaxy clustering, the $32$ galaxy-shear pairs for which the mean lens redshift is lower than the mean source redshift, and all $15$ shear-shear pairs ($5$ auto- and $10$ cross-correlations). Combined with the $24$ bandpowers described above, the resulting data vector has $(10 + 32 + 15) \times 24 = 1368$ elements.
All inference is performed within the Julia \texttt{Turing.jl}\footnote{\url{https://github.com/TuringLang/Turing.jl}} framework \citep{fjelde2025turing}, with gradients of the \blast{} likelihood obtained through reverse-mode automatic differentiation via \texttt{Mooncake.jl}\footnote{\url{https://github.com/chalk-lab/Mooncake.jl}}.

As a first test, we sample only the $5$ cosmological parameters, which allows us to compare three samplers: the gradient-free \texttt{emcee} \citep{foreman2013emcee}, the No-U-Turn Sampler (\texttt{NUTS}) \citep{hoffman2014no}, and the Microcanonical Langevin Monte Carlo sampler (\texttt{MCLMC}) \citep{robnik2026metropolis}. For \texttt{NUTS} and \texttt{MCLMC}, chains are initialized using \texttt{PathFinder.jl}\footnote{\url{https://mlcolab.github.io/Pathfinder.jl/stable/}} \citep{zhang2022pathfinder}, which provides a variational approximation to the posterior as a starting point for the chains, significantly reducing the warm-up time required to reach the typical set.

The prior distributions for the five sampled cosmological parameters are listed in Table~\ref{tab:priors}. All priors are substantially wider than the resulting posteriors and are uninformative for every parameter shown.

\begin{table}[h!]
\centering
\caption{Prior distributions for the five sampled cosmological parameters. $\mathcal{U}(a,b)$ denotes a uniform prior on $[a,b]$; $\mathcal{N}(\mu,\sigma)$ denotes a Gaussian prior with mean $\mu$ and standard deviation $\sigma$, truncated to the stated range.}
\label{tab:priors}
\begin{tabular}{lll}
\hline\hline
Parameter & Prior & Range \\
\hline
$h$ & $\mathcal{U}(0.64, 0.73)$ & $[0.64, 0.73]$ \\
$\omega_\mathrm{b}$ & $\mathcal{U}(0.020, 0.025)$ & $[0.020, 0.025]$ \\
$\Omega_\mathrm{m}$ & $\mathcal{U}(0.27, 0.34)$ & $[0.27, 0.34]$ \\
$\ln(10^{10}A_s)$ & $\mathcal{N}(3.054, 0.14)$ & $[2.0, 3.5]$ \\
$n_s$ & $\mathcal{N}(0.9645, 0.015)$ & $[0.80, 1.10]$ \\
\hline
\end{tabular}
\end{table}

For \texttt{emcee}, we use 20 walkers with $20,000$ steps each, discarding the first $25\%$ as burn-in, yielding $300,000$ pooled post-burn-in samples. For \texttt{NUTS}, we run 5 parallel chains with $1,000$ post-warmup samples per chain and target acceptance $0.85$, yielding $5{,}000$ pooled samples. For \texttt{MCLMC}, we run 5 independent chains with $50,000$ post-adaptation samples each, using a three-phase self-tuning adaptation of $5,000$ steps targeting an energy-error variance of $5\cdot10^{-5}$, yielding $200,000$ pooled samples.

\begin{figure}[h!]
    \centering\includegraphics[width=0.97\columnwidth]{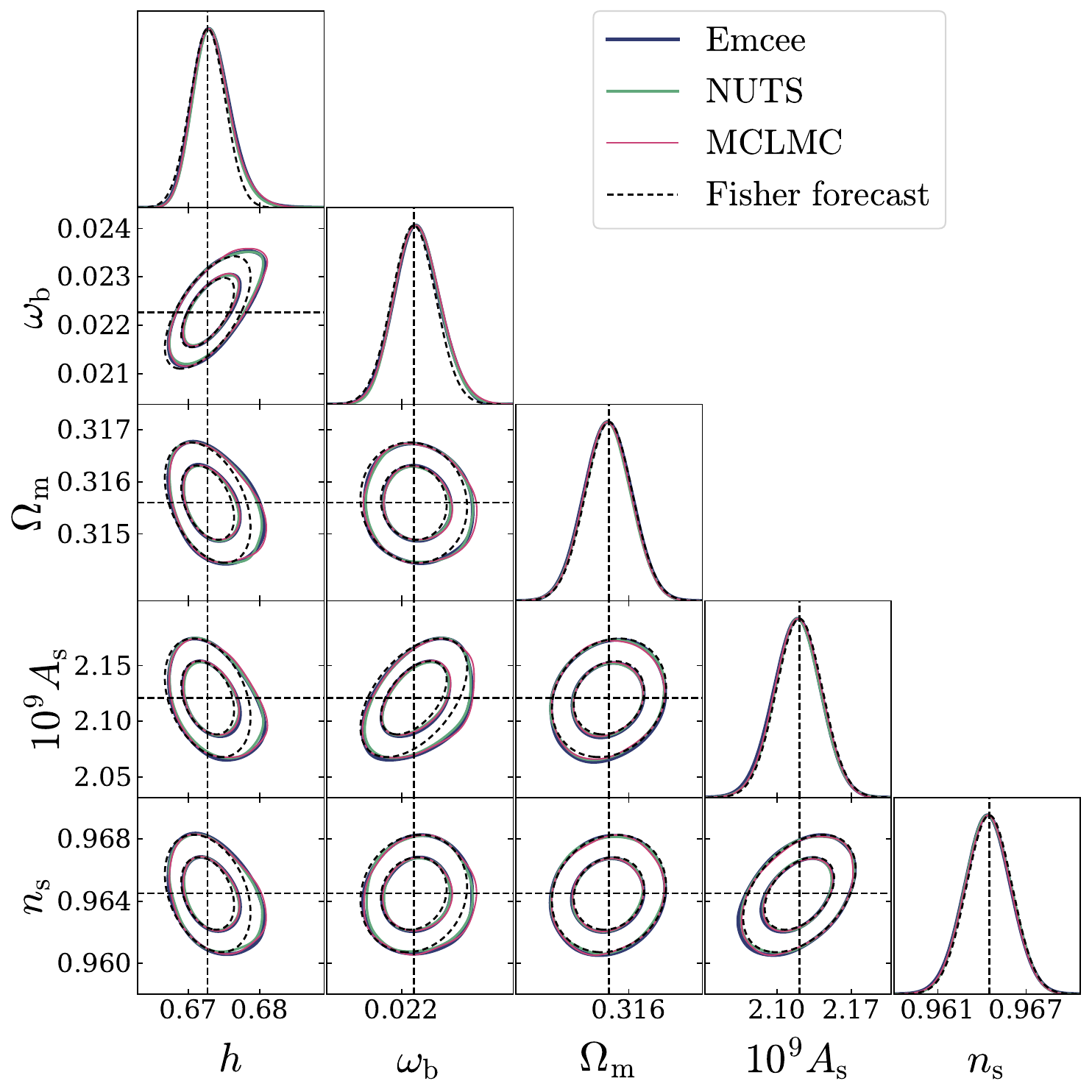}
    \caption{Marginalized posterior distributions for the five cosmological parameters from the cosmology-only ($5$-parameter) inference. Three samplers are compared: \texttt{emcee} ($300{,}000$ samples), \texttt{NUTS} ($5{,}000$ samples), and \texttt{MCLMC} ($200{,}000$ samples). A Fisher forecast is shown for reference (dashed contours). Dotted lines indicate the fiducial parameter values. All three samplers are in excellent mutual agreement.}
    \label{fig:chains_cosmo}
\end{figure}

Convergence was assessed via split-$\hat{R}$ and effective sample size; across all three samplers and all five parameters, $\hat{R} \le 1.013$ and the effective sample size ranges from $\sim$590 to $\sim$4,400. The marginalized posteriors are shown in Fig.~\ref{fig:chains_cosmo}, together with a Fisher forecast for reference. All three samplers are in excellent mutual agreement and recover the fiducial cosmology within their uncertainties.

All chains were run on single AMD EPYC Genoa compute node on the Narval cluster using 32 CPUs. \texttt{NUTS} completed in approximately $3$ hours per chain, \texttt{MCLMC} in approximately $20$ hours per chain, and \texttt{emcee} in approximately $16$ hours in total.

We then extend the analysis to the complete set of nuisance parameters alongside cosmology, for a total of $35$ free parameters: $5$ cosmological parameters, $10$ linear galaxy bias amplitudes (one per clustering bin), $10$ redshift-distribution shift parameters (one per clustering bin), $5$ multiplicative shear bias parameters (one per source bin), and $5$ intrinsic alignment amplitudes (one per source bin). Since \texttt{emcee}'s gradient-free proposal does not scale to this dimensionality, we use only \texttt{NUTS} and \texttt{MCLMC}, with the same sampler configurations and cosmological priors as in the cosmology-only case. The nuisance parameters are assigned wide uniform priors as listed in Table~\ref{tab:priors_nuisance}, all substantially wider than the resulting posteriors.

\begin{table}[h!]
\centering
\caption{Prior distributions for the $30$ nuisance parameters in the full $35$-parameter inference. $\mathcal{U}(a,b)$ denotes a uniform prior on $[a,b]$. All priors are substantially wider than the resulting posteriors.}
\label{tab:priors_nuisance}
\begin{tabular}{lll}
\hline\hline
Parameter & Prior & Description \\
\hline
$b_i$ & $\mathcal{U}(0.3, 3.5)$ & Galaxy bias \\
$\Delta z_i$ & $\mathcal{U}(-0.05, 0.05)$ & $n(z)$ shift \\
$m_i$ & $\mathcal{U}(-0.1, 0.1)$ & Mult. Shear bias \\
$A_{{\rm IA},i}$ & $\mathcal{U}(-5.0, 5.0)$ & IA amplitude \\
\hline
\end{tabular}
\end{table}

As mentioned, \texttt{NUTS} and \texttt{MCLMC} use the same \texttt{PathFinder.jl} initialization, with maximum tree depth capped at $6$ and target acceptance $0.85$ for \texttt{NUTS} and a target energy of $5\cdot10^{-5}$ for \texttt{MCLMC} as before.

The marginalized posteriors for the five cosmological parameters, after marginalizing over all $30$ nuisance parameters, are shown in Fig.~\ref{fig:chains_full}. The nuisance-parameter posteriors from the same chains are presented in Appendix~\ref{app:corner_plot}.

\begin{figure}[h!]
    \centering\includegraphics[width=0.97\columnwidth]{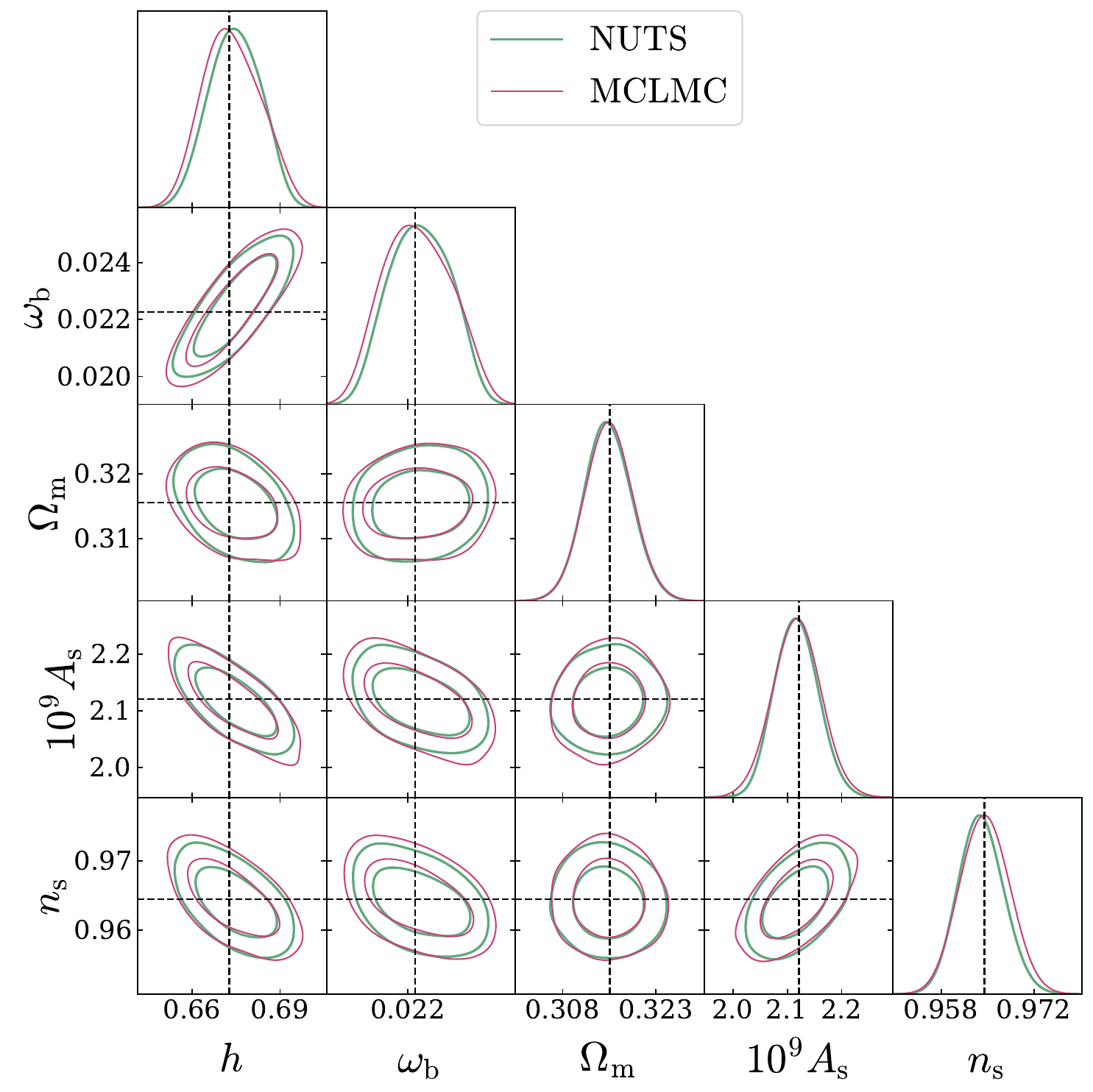}
    \caption{Marginalized posterior distributions for the five cosmological parameters from the full $35$-parameter inference, after marginalizing over the $30$ nuisance parameters. Results from \texttt{NUTS} and \texttt{MCLMC} are shown; both samplers are in excellent mutual agreement and recover the fiducial input cosmology (dotted lines) within their respective uncertainties. The corresponding nuisance-parameter posteriors are shown in Appendix~\ref{app:corner_plot}.}
    \label{fig:chains_full}
\end{figure}

For the full 35-parameter inference, we ran the 15 chains for each sampler to reach convergence, consistent with the harder posterior geometry expected at seven times the dimensionality of the cosmology-only case. This yields $15 \times 1{,}000 = 15{,}000$ pooled post-warmup samples for \texttt{NUTS} and $15 \times 50{,}000 = 750{,}000$ pooled post-adaptation samples for \texttt{MCLMC}. 
\texttt{NUTS} chains completed in approximately $17$ hours each, and \texttt{MCLMC} chains in approximately $24$ hours each, using the same single-node, $32$-CPU configuration as the cosmology-only run.

Beyond the per-chain wall-clock cost reported above, the most insightful benchmark is the cost of a model evaluation, since this is the fundamental unit repeated at every step of the inference. The timings reported in Sec.~\ref{sec:performance} and Table~\ref{tab:fid_res} characterize the \blast{} $C_\ell$ computation in the \nk{} challenge configuration; here we report the cost of the complete inference pipeline: from the cosmological parameters, through the power spectrum evaluation with \texttt{Mapse.jl}, to $C_\ell$'s obtained with the \blast{} algorithm and finally the evaluation of the log-likelihood. In addition, we report the cost of the reverse-mode \texttt{Mooncake.jl} gradient required at each leapfrog step. Beyond the $5$ physical cosmological parameters, we also stress-test the scaling of the reverse-mode gradient by independently differentiating every entry of the tabulated power spectra that \blast{} operates on internally: the $50\times161$ non-limber grid plus the two $50\times257$ limber (linear and nonlinear) grids, for a total of $33{,}750$ independent inputs. 
Timings were obtained with \texttt{BenchmarkTools.jl}  on both AMD EPYC Rome CPUs (for direct comparison with the rest of the paper and \citep{chiarenza2024blast}) and AMD EPYC Genoa nodes, the newer and better performing CPUs available on the Narval cluster. As this architecture achieves its best performance at $32$ threads, we fix the thread count to $32$ throughout. To mitigate node-to-node performance variance on the cluster, each timing was evaluated across ten identical jobs submitted to different nodes, with exclusive node access requested to reduce timing noise, and the minimum median runtime across the ten jobs was retained.

\begin{table}[h!]
\centering
\begin{tabular}{lrrr}
\hline
Configuration & Fwd (ms) & Grad (ms) & Total (ms) \\
\hline
Rome, 5-param  & 145 & 483  &  628 \\
Rome, 33750-param & 131 & 496 & 627 \\
Genoa, 5-param &  42 & 251 & 293 \\
Genoa, 33750-param &  45 & 247 & 292 \\
\hline
\end{tabular}
\caption{Median wall-clock cost of a single full-model evaluation for the $5$- and $33{,}750$-parameter cases. \emph{Fwd}: log-likelihood evaluation only. \emph{Grad}: \texttt{Mooncake.jl} reverse-mode gradient. \emph{Total} is effectively the cost of each inference step.}
\label{tab:full_model_timing}
\end{table}

Table~\ref{tab:full_model_timing} shows how the cost of the gradient evaluation is nearly identical whether differentiating $5$ physical parameters or all $33{,}750$ entries of the tabulated power spectra. This is the expected signature of reverse-mode automatic differentiation, whose cost scales with the size of the computational graph rather than the number of differentiated inputs, and it holds even as the number of differentiated inputs grows by almost four orders of magnitude. Second, Genoa nodes deliver a $\sim2\times$ speedup over Rome because of the newer architecture, motivating our choice of using this hardware for the chains.

\section{Conclusions}\label{sec:conclusions}
In this work, we presented a major extension of \blast{}, transforming it from the version tailored to the \nk{} challenge \citep{chiarenza2024blast} into a comprehensive, differentiable toolkit for full physics cosmological analyses. Building on the Chebyshev polynomial decomposition that was defined in the original framework, we generalized the treatment of the matter power spectrum to a transfer-function formulation, enabling a consistent and efficient treatment of primordial non-Gaussianity, redshift-space distortions, magnification bias, and intrinsic alignments, alongside cross-correlations with the cosmic microwave background through CMB lensing and the integrated Sachs-Wolfe effect.

These extensions required a substantial generalization of the underlying numerical infrastructure: the introduction of multiple power spectrum classes to capture scale-dependent bias, the inclusion of second derivatives of spherical Bessel functions for RSD, and an expanded set of $17$ projected matter densities from which all auto- and cross-correlations are assembled. Despite this increase in physical complexity, \blast{} retains the favorable computational scaling of the original algorithm by pre-computing all cosmology-independent quantities, and becomes fully differentiable through custom automatic differentiation rules.

We validated the extended framework extensively, demonstrating sub-percent agreement against both a high-precision brute-force integration and a broad suite of established cosmological codes, including \texttt{CAMB}, \texttt{CLASS}, \texttt{CCL}, \texttt{SwiftC}$_\ell$, and \texttt{pylevin}, across all supported physical effects. We further showed that \blast{} remains the most computationally efficient non-Limber method among those benchmarked, running in milliseconds even with the full set of effects enabled. Our analysis of the algorithm's hyperparameters showed that accuracy and runtime can be jointly tuned via $\mathrm{n}\chi$, $\mathrm{n}R$, and $n_{\rm cheb}$ parameters, and we provided a default configuration, together with precomputed basis integrals, intended to be accurate enough for most realistic survey analyses.

Finally, we demonstrated the readiness of \blast{} for modern, gradient-based cosmological inference through a simulated likelihood analysis using LSST Y10-like synthetic data, sampling the posterior with the No-U-Turn Sampler and Microcanonical Langevin Monte Carlo sampler via reverse-mode automatic differentiation. Both samplers recover the fiducial cosmology within their uncertainties across two inference scenarios: a cosmology-only ($5$-parameter) analysis benchmarked also against the gradient-free sampler \texttt{emcee}, and a full $35$-parameter analysis. Notably, the cost of a full forward-plus-gradient evaluation is nearly identical for both models ($\sim290\,\mathrm{ms}$ on modern hardware), a direct consequence of reverse-mode automatic differentiation whose cost scales with the computational graph rather than the number of free parameters.
Finally, a GPU-compatible version of \blast{} is currently under development. The GPU implementation requires no fundamental changes to the algorithm, only targeted adjustments to ensure efficient execution on GPU architectures, and is expected to deliver a further speedup over the already competitive CPU runtimes reported here. We will present and release this extension in forthcoming work applying \blast{} to cosmological analyses. As gradient-based inference on GPU hardware becomes increasingly standard in the field, this extension will ensure that \blast{} remains well-suited to modern high-performance computing environments.

Beyond this validation and forecasting effort, \blast{} has already been applied to real survey data: it was used to model the scale-dependent bias signature of local-type primordial non-Gaussianity in the cross-correlation between $1.2$ million DESI DR1 quasars and \textit{Planck} PR4 CMB lensing \citep{chiarenza2025constraining}. A follow-up analysis extending this measurement to the larger DESI DR2 samples is currently in preparation \citep{chiarenza2026inprep}.

The growing interest of the cosmological community in gradient-based inference algorithms \citep{campagne2023jax, piras2023cosmopower, balkenhol2024candl, ruiz2023limberjack, hahn2024disco} motivates the development of fast, differentiable theory pipelines such as \blast{}. Recent applications have already demonstrated the power of these methods in realistic cosmological analyses \citep{zhang2025enhancing, bonici2026alleviating}. The availability of gradients also opens the door to frequentist analyses based on the profile likelihood \citep{herold2025profile, morawetz2025frequentist}, which offer a complementary perspective to the Bayesian posteriors presented here and are particularly valuable when the posterior distribution is non-Gaussian (e.g., in the presence of asymmetric tails or multi-modal structure) where Bayesian credible intervals and frequentist confidence intervals can differ substantially.

Looking further ahead, the modular \julia{} architecture of \blast{} makes it naturally compatible with other \julia{}-based cosmological codes, enabling joint analyses that extend well beyond the multi-probe framework considered here. In particular, it will be possible to combine \blast{} with CMB data pipelines \citep{bonici2023capse}, spectroscopic galaxy clustering multipoles \citep{bonici2025effort}, and eventually field-level analyses \citep{loureiro2022almanac, crespi2025flinch}. Finally, \blast{} is currently being integrated into the official Euclid analysis pipeline \citep{bonici2026cloelib}, where it will contribute to forthcoming Euclid cosmological analyses, a concrete demonstration of its readiness for next-generation survey science.

\section*{Acknowledgments}
The authors wish to thank Robert Reischke, Laura Reymond, and Alexander Reeves for helpful discussions and assistance with the \texttt{pylevin} and \swift{} codes, respectively.  We also thank Anthony Lewis and Vivian Miranda for useful discussions, and Noah Sailer for providing independent brute-force angular power spectra that served as an additional validation of \blast{}.
The authors acknowledge the support of the Canadian Space Agency. WP also acknowledges support from the Natural Sciences and Engineering Research Council of Canada (NSERC), [funding reference number RGPIN-2025-03931]. CGG is supported by the C\'esar Nombela  Research Talent Attraction grant from the Community of Madrid  (Ref. 2025-T1/TEC-36302).
Research at Perimeter Institute is supported in part by the Government of Canada through the Department of Innovation, Science and Economic Development Canada and by the Province of Ontario through the Ministry of Colleges and Universities. This research was enabled in part by support provided by Compute Ontario (computeontario.ca) and the Digital Research Alliance of Canada (alliancecan.ca).

\appendix
\section{Summary of the building blocks}\label{app:contributions}
\begin{table*}[h!]
\centering
\renewcommand{\arraystretch}{1.5}
\centering
\caption{Detailed summary of the constituent physical components for the angular power spectra implemented in \blast{}. For each contribution, we specify the multipole-dependent prefactor, the $k$-scaling of the integrand, the configuration of the spherical Bessel functions, and the required matter power spectrum combination. The \texttt{w\_ell} ID identifies the specific projected matter density computed within the \blast{} framework. These individual elements serve as the fundamental building blocks for constructing the full auto- and cross-correlation observables.}
\label{tab:w_ell_summary}
\begin{tabular}{l l c c c l}
\toprule
Contribution & $\ell$ prefactor & $k$-scaling & Bessel functions & Power spectrum & Projected matter density $w_\ell$ \\
\midrule
$C_\ell^{\delta\delta}$ & $1$ & $k^2$ & $j_\ell, j_\ell$ & $P_\Phi T_\mathrm{m}^2$ & \texttt{w\_2\_00\_fTT} \\
$C_\ell^{\delta\mu}$ & $\ell(\ell+1)$ & $1$ & $j_\ell, j_\ell$ & $P_\Phi T_\mathrm{m}^2$ & \texttt{w\_0\_00\_fTT} \\
$C_\ell^{\delta-\mathrm{RSD}}$ & $1$ & $k^2$ & $j_\ell, j^{''}_\ell$ & $P_\Phi T_\mathrm{m}^2$ & \texttt{w\_2\_02\_fTT}, \texttt{w\_2\_20\_fTT} \\
$C_\ell^{\mu-\mathrm{RSD}}$ & $\ell(\ell+1)$ & $1$ & $j_\ell, j^{''}_\ell$ & $P_\Phi T_\mathrm{m}^2$ & \texttt{w\_0\_02\_fTT}, \texttt{w\_0\_20\_fTT} \\
$C_\ell^{\mu\mu}$ & $(\ell(\ell+1))^2$ & $1/k^2$ & $j_\ell, j_\ell$ & $P_\Phi T_\mathrm{m}^2$ & \texttt{w\_-2\_00\_fTT} \\
$C_\ell^{\mathrm{RSD}-\mathrm{RSD}}$ & $1$ & $k^2$ & $j^{''}_\ell, j^{''}_\ell$ & $P_\Phi T_\mathrm{m}^2$ & \texttt{w\_2\_22\_fTT} \\
$C_\ell^{\mu-f_\mathrm{NL}}$ & $\ell(\ell+1)$ & $k^2$ & $j_\ell, j_\ell$ & $P_\Phi T_\mathrm{m}$ & \texttt{w\_0\_00\_fT} \\
$C_\ell^{f_\mathrm{NL}-\mathrm{RSD}}$ & $1$ & $k^2$ & $j_\ell, j^{''}_\ell$ & $P_\Phi T_\mathrm{m}^2$ & \texttt{w\_2\_02\_fT}, \texttt{w\_2\_20\_fT} \\
$C_\ell^{\delta-f_\mathrm{NL}}$ & $1$ & $k^2$ & $j_\ell, j_\ell$ & $P_\Phi T_\mathrm{m}$ & \texttt{w\_2\_00\_fT} \\
$C_\ell^{f_\mathrm{NL}-f_\mathrm{NL}}$ & $1$ & $k^2$ & $j_\ell, j_\ell$ & $P_\Phi$ & \texttt{w\_2\_00\_f} \\
$C_\ell^{\delta\gamma}$ & $\sqrt{\frac{(\ell+2)!}{(\ell-2)!}}$ & $1$ & $j_\ell, j_\ell$ & $P_\Phi T_\mathrm{m}^2$ & \texttt{w\_0\_00\_fTT} \\
$C_\ell^{\delta-\mathrm{IA}}$ & $\sqrt{\frac{(\ell+2)!}{(\ell-2)!}}$ & $1$ & $j_\ell, j_\ell$ & $P_\Phi T_\mathrm{m}^2$ & \texttt{w\_0\_00\_fTT} \\
$C_\ell^{\mathrm{RSD}-\gamma}$ & $\sqrt{\frac{(\ell+2)!}{(\ell-2)!}}$ & $1$ & $j^{''}_\ell, j_\ell$ & $P_\Phi T_\mathrm{m}^2$ & \texttt{w\_0\_20\_fTT}, \texttt{w\_0\_02\_fTT} \\
$C_\ell^{\mathrm{RSD}-\mathrm{IA}}$ & $\sqrt{\frac{(\ell+2)!}{(\ell-2)!}}$ & $1$ & $j^{''}_\ell, j_\ell$ & $P_\Phi T_\mathrm{m}^2$ & \texttt{w\_0\_20\_fTT}, \texttt{w\_0\_02\_fTT} \\
$C_\ell^{\mu\gamma}$ & $\ell(\ell+1)\sqrt{\frac{(\ell+2)!}{(\ell-2)!}}$ & $1/k^2$ & $j_\ell, j_\ell$ & $P_\Phi T_\mathrm{m}^2$ & \texttt{w\_-2\_00\_fTT} \\
$C_\ell^{\mu-\mathrm{IA}}$ & $\ell(\ell+1)\sqrt{\frac{(\ell+2)!}{(\ell-2)!}}$ & $1/k^2$ & $j_\ell, j_\ell$ & $P_\Phi T_\mathrm{m}^2$ & \texttt{w\_-2\_00\_fTT} \\
$C_\ell^{f_\mathrm{NL}-\gamma}$ & $\sqrt{\frac{(\ell+2)!}{(\ell-2)!}}$ & $1$ & $j_\ell, j_\ell$ & $P_\Phi T_\mathrm{m}$ & \texttt{w\_0\_00\_fT} \\
$C_\ell^{f_\mathrm{NL}-\mathrm{IA}}$ & $\sqrt{\frac{(\ell+2)!}{(\ell-2)!}}$ & $1$ & $j_\ell, j_\ell$ & $P_\Phi T_\mathrm{m}$ & \texttt{w\_0\_00\_fT} \\
$C_\ell^{\delta\kappa}$ & $\ell(\ell+1)$ & $1$ & $j_\ell, j_\ell$ & $P_\Phi T_\mathrm{m}^2$ & \texttt{w\_0\_00\_fTT} \\
$C_\ell^{\delta T}$ & $1$ & $k^2$ & $j_\ell, j_\ell$ & $P_\Phi T_\mathrm{m}^2$ & \texttt{w\_0\_00\_fTT} \\
$C_\ell^{\mathrm{RSD}-\kappa}$ & $\ell(\ell+1)$ & $1$ & $j^{''}_\ell, j_\ell$ & $P_\Phi T_\mathrm{m}^2$ & \texttt{w\_0\_02\_fTT}, \texttt{w\_0\_20\_fTT} \\
$C_\ell^{\mathrm{RSD}-T}$ & $1$ & $1$ & $j^{''}_\ell, j_\ell$ & $P_\Phi T_\mathrm{m}^2$ & \texttt{w\_0\_02\_fTT}, \texttt{w\_0\_20\_fTT} \\
$C_\ell^{\kappa\mu}$ & $(\ell(\ell+1))^2$ & $1/k^2$ & $j_\ell, j_\ell$ & $P_\Phi T_\mathrm{m}^2$ & \texttt{w\_-2\_00\_fTT} \\
$C_\ell^{\mu T}$ & $\ell(\ell+1)$ & $1/k^2$ & $j_\ell, j_\ell$ & $P_\Phi T_\mathrm{m}^2$ & \texttt{w\_-2\_00\_fTT} \\
$C_\ell^{\kappa-f_\mathrm{NL}}$ & $\ell(\ell+1)$ & $1$ & $j_\ell, j_\ell$ & $P_\Phi T_\mathrm{m}$ & \texttt{w\_0\_00\_fT} \\
$C_\ell^{T-f_\mathrm{NL}}$ & $1$ & $1$ & $j_\ell, j_\ell$ & $P_\Phi T_\mathrm{m}$ & \texttt{w\_0\_00\_fT} \\
$C_\ell^{\kappa\gamma}$ & $\ell(\ell+1)\sqrt{\frac{(\ell+2)!}{(\ell-2)!}}$ & $1/k^2$ & $j_\ell, j_\ell$ & $P_\Phi T_\mathrm{m}^2$ & \texttt{w\_-2\_00\_fTT} \\
$C_\ell^{\kappa-\mathrm{IA}}$ & $\ell(\ell+1)\sqrt{\frac{(\ell+2)!}{(\ell-2)!}}$ & $1/k^2$ & $j_\ell, j_\ell$ & $P_\Phi T_\mathrm{m}^2$ & \texttt{w\_-2\_00\_fTT} \\
$C_\ell^{\gamma T}$ & $\sqrt{\frac{(\ell+2)!}{(\ell-2)!}}$ & $1/k^2$ & $j_\ell, j_\ell$ & $P_\Phi T_\mathrm{m}^2$ & \texttt{w\_-2\_00\_fTT} \\
$C_\ell^{T-\mathrm{IA}}$ & $\sqrt{\frac{(\ell+2)!}{(\ell-2)!}}$ & $1/k^2$ & $j_\ell, j_\ell$ & $P_\Phi T_\mathrm{m}^2$ & \texttt{w\_-2\_00\_fTT} \\
$C_\ell^{\gamma\gamma}$ & $\frac{(\ell+2)!}{(\ell-2)!}$ & $1/k^2$ & $j_\ell, j_\ell$ & $P_\Phi T_\mathrm{m}^2$ & \texttt{w\_-2\_00\_fTT} \\
$C_\ell^{\gamma-\mathrm{IA}}$ & $\frac{(\ell+2)!}{(\ell-2)!}$ & $1/k^2$ & $j_\ell, j_\ell$ & $P_\Phi T_\mathrm{m}^2$ & \texttt{w\_-2\_00\_fTT} \\
$C_\ell^{\mathrm{IA}-\mathrm{IA}}$ & $\frac{(\ell+2)!}{(\ell-2)!}$ & $1/k^2$ & $j_\ell, j_\ell$ & $P_\Phi T_\mathrm{m}^2$ & \texttt{w\_-2\_00\_fTT} \\
\bottomrule
\end{tabular}
\end{table*}

The \blast{} framework adopts a modular architecture for the computation of angular power spectra. As described in Sec.~\ref{sec:extensions}, we decompose the observables into their constituent physical components. For the joint multi-probe analyses considered in this work, the primary observed fields—galaxy clustering ($\mathrm{g}$), weak lensing ($\epsilon$), and CMB ($\mathrm{T}$)—can be schematically represented as:
\begin{align}
    \mathrm{g} &= \delta + \mathrm{RSD} + \mu + \mathrm{PNG} \;, \\
    \mathrm{s} &= \gamma + \mathrm{IA} \;, \\
    \mathrm{T} &= \kappa + \mathrm{ISW} \;,
\end{align}
where the individual terms correspond to number counts, redshift-space distortions, magnification bias, primordial non-Gaussianity, cosmic shear, intrinsic alignments, CMB lensing, and the integrated Sachs-Wolfe effect, respectively.

Under this decomposition, any angular power spectrum $C_\ell^{\mathrm{XY}}$ between fields $\mathrm{X}$ and $\mathrm{Y}$ is constructed as the sum of the cross-correlations of their individual components. Table~\ref{tab:w_ell_summary} summarizes the specific kernels, prefactors, and projected matter density ($w_\ell$) used to build these signals.

We also note that, in the general case with tomographic binning, the $C_\ell$'s are not symmetric in the components: $C_{i,j}^{\mathrm{AB}} \neq C_{ i,j}^{\mathrm{BA}}$. In general, it is true that $C_{i,j}^{\mathrm{AB}}= C_{j,i}^{\mathrm{BA}}$, but in our case we cannot exploit this symmetry in a useful way because of the change of variables to $\chi - R$ (see Sec.~\ref{sec:original_blast}).

By combining the components listed in Table~\ref{tab:w_ell_summary}, we can express the primary observables. The galaxy clustering auto-spectrum, being the most complex case, expands into 16 individual terms:

\begin{align}
C_\ell^{\mathrm{gg}}
&= C_\ell^{(\delta+\mathrm{RSD}+\mu+f_\mathrm{NL})(\delta+\mathrm{RSD}+\mu+f_\mathrm{NL})}\nonumber \\
&= C_\ell^{\delta\delta}
 + C_\ell^{\delta\mu}
 + C_\ell^{\delta-\mathrm{RSD}}
 + C_\ell^{\delta-f_\mathrm{NL}}\nonumber \\
&\quad +\, C_\ell^{\mu\delta}
 + C_\ell^{\mu\mu}
 + C_\ell^{\mu-\mathrm{RSD}}
 + C_\ell^{\mu-f_\mathrm{NL}}\nonumber \\
&\quad +\, C_\ell^{\mathrm{RSD}-\delta}
 + C_\ell^{\mathrm{RSD}-\mu}
 + C_\ell^{\mathrm{RSD}-\mathrm{RSD}}
 + C_\ell^{\mathrm{RSD}-f_\mathrm{NL}}\nonumber \\
&\quad +\, C_\ell^{f_\mathrm{NL}-\delta}
 + C_\ell^{f_\mathrm{NL}-\mu}
 + C_\ell^{f_\mathrm{NL}-\mathrm{RSD}}
 + C_\ell^{f_\mathrm{NL}-f_\mathrm{NL}}\;.
\end{align}

The remaining auto- and cross-probe observables are constructed similarly:

\paragraph{Weak lensing}
\begin{equation}
\begin{alignedat}{1}
C_\ell^{\mathrm{ss}}
&= C_\ell^{(\gamma+\mathrm{IA})(\gamma+\mathrm{IA})} \\[4pt]
&= C_\ell^{\gamma\gamma}
 + C_\ell^{\gamma-\mathrm{IA}}
 + C_\ell^{\mathrm{IA}\gamma}
 + C_\ell^{\mathrm{IA}-\mathrm{IA}} .
\end{alignedat}
\end{equation}

\paragraph{Galaxy Clustering $\times$ Weak Lensing}

\begin{equation}
\begin{alignedat}{1}
C_\ell^{\mathrm{gs}}
&= C_\ell^{(\delta+\mathrm{RSD}+\mu+f_\mathrm{NL})(\gamma+\mathrm{IA})} \\[4pt]
&= C_\ell^{\delta\gamma}
 + C_\ell^{\delta-\mathrm{IA}}
 + C_\ell^{\mathrm{RSD}-\gamma}
 + C_\ell^{\mathrm{RSD}-\mathrm{IA}} \\[4pt]
&\quad +\, C_\ell^{\mu\gamma}
 + C_\ell^{\mu-\mathrm{IA}}
 + C_\ell^{f_\mathrm{NL}-\gamma}
 + C_\ell^{f_\mathrm{NL}-\mathrm{IA}} .
\end{alignedat}
\end{equation}

\paragraph{Galaxy Clustering $\times$ CMB}
\begin{equation}
\begin{alignedat}{1}
C_\ell^{\mathrm{gT}}
&= C_\ell^{(\delta+\mathrm{RSD}+\mu+f_\mathrm{NL})(\kappa+\mathrm{ISW})} \\[4pt]
&= C_\ell^{\delta\kappa}
 + C_\ell^{\delta-\mathrm{ISW}}
 + C_\ell^{\mathrm{RSD}-\kappa}
 + C_\ell^{\mathrm{RSD}-\mathrm{ISW}} \\[4pt]
&\quad +\, C_\ell^{\mu\kappa}
 + C_\ell^{\mu-\mathrm{ISW}}
 + C_\ell^{f_\mathrm{NL}-\kappa}
 + C_\ell^{f_\mathrm{NL}-\mathrm{ISW}} .
\end{alignedat}
\end{equation}

\paragraph{Weak Lensing $\times$ CMB}
\begin{equation}
\begin{alignedat}{1}
C_\ell^{\mathrm{sT}}
&= C_\ell^{(\gamma+\mathrm{IA})(\kappa+\mathrm{ISW})} \\[4pt]
&= C_\ell^{\gamma\kappa}
 + C_\ell^{\gamma-\mathrm{ISW}}
 + C_\ell^{\mathrm{IA}-\kappa}
 + C_\ell^{\mathrm{IA}-\mathrm{ISW}} .
\end{alignedat}
\end{equation}

\newpage
\section{Expanded comparison plot}\label{app:master_comparison}
\begin{figure}[h!]
    \centering
    \includegraphics[width=0.73\textwidth]{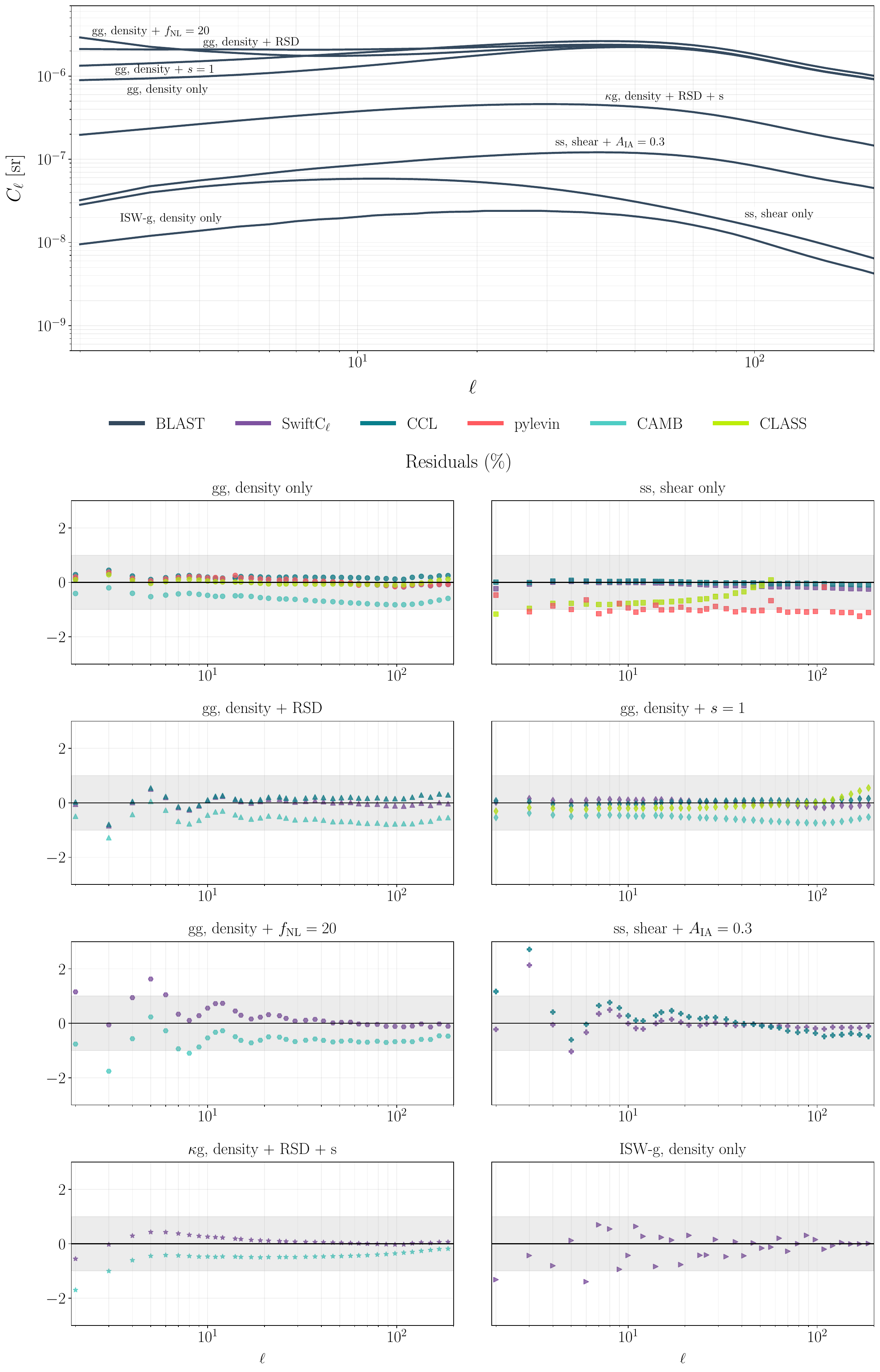}
    \caption{Detailed breakdown of residuals for the multi-code comparison. This figure replicates the benchmark analysis presented in Fig.~\ref{fig:master_plot}, but organizes the percentage residuals into individual subpanels for each physical configuration to provide a clearer view of specific code agreements. The top panel displays the reference \blast{} $C_\ell$ spectra, while the bottom grid shows the residuals relative to \blast{} for each corresponding auto- and cross-probe. Gray shaded regions indicate the $\pm 1\%$ residual threshold.}
    \label{fig:master_comparison_exp}
\end{figure}

\section{Full nuisance parameters corner plot}\label{app:corner_plot}
\begin{figure}[h!]
    \centering
    \includegraphics[width=\textwidth]{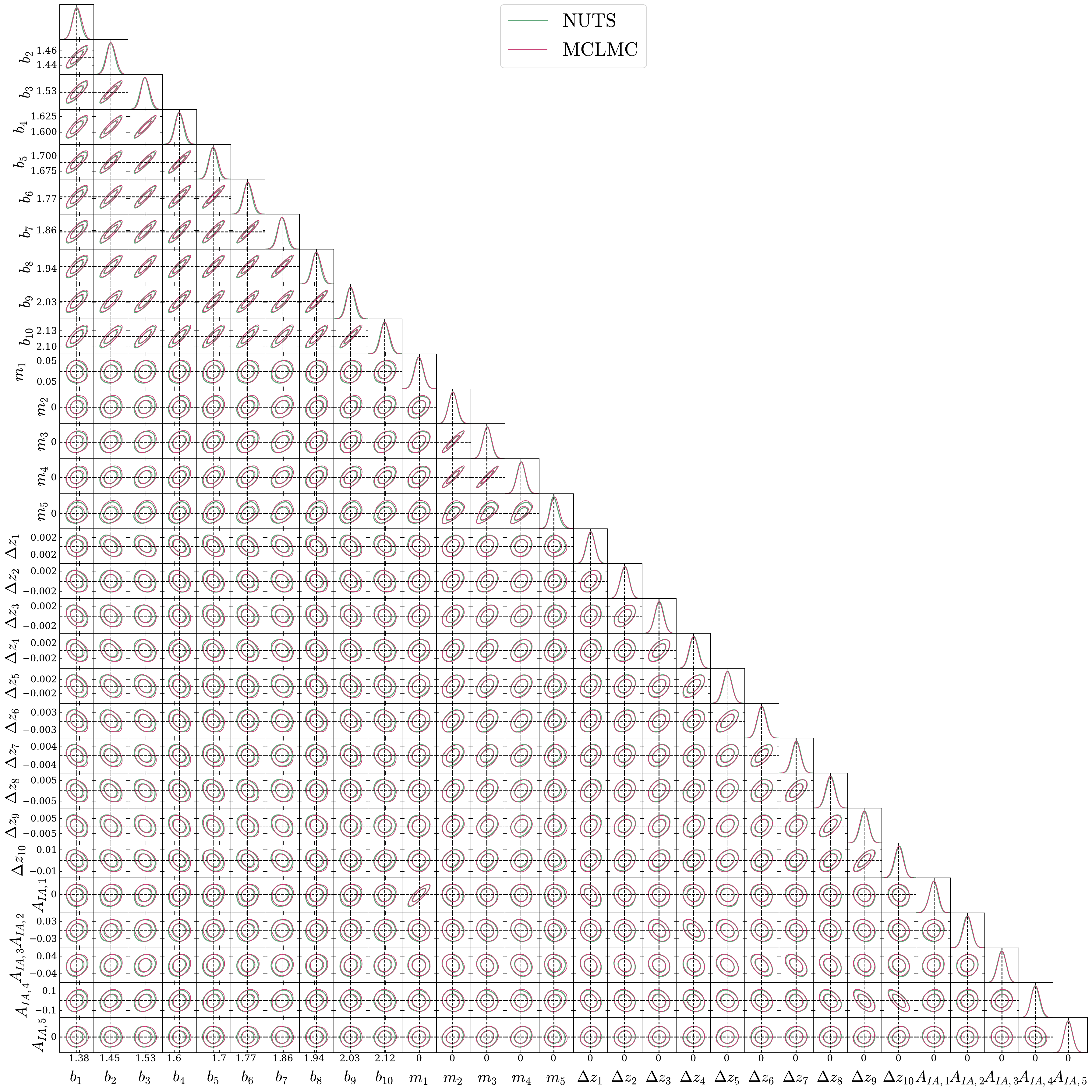}
    \caption{As Fig.~\ref{fig:chains_full}, but for the $30$ nuisance parameters: galaxy bias, multiplicative shear bias, photo-z shift, and intrinsic-alignment amplitude.}
    \label{fig:nuisances}
\end{figure}

\bibliographystyle{apsrev4-2}
\bibliography{biblio}
\end{document}